\documentclass[structabstract]{aa} 
\usepackage[varg]{txfonts}
\usepackage{graphicx}
\usepackage{longtable,lscape}

\begin{document}

   \title{Deep SDSS optical spectroscopy of distant halo stars}

   \subtitle{II. Iron, calcium, and magnesium abundances.}
   \author{E. Fern\'andez-Alvar\inst{1,2}
          \and
          C. Allende Prieto\inst{1,2}
      \and
          K. J. Schlesinger\inst{3}
      \and
          T. C. Beers\inst{4} 
      \and 
          A. C. Robin\inst{5}      
      \and
          D. P. Schneider\inst{6,7}       
      \and
          Y. S. Lee\inst{8}          
      \and
         D. Bizyaev\inst{9}
      \and
         G. Ebelke\inst{9}
      \and
         E. Malanushenko\inst{9}
      \and
         V. Malanushenko\inst{9}
      \and
         D. Oravetz\inst{9}
      \and
          K. Pan\inst{9}
      \and
          A. Simmons\inst{9}
        }

   \institute{Instituto de Astrof\'{\i}sica de Canarias,
              V\'{\i}a L\'actea, 38205 La Laguna, Tenerife, Spain \\             
         \and
             Universidad de La Laguna, Departamento de Astrof\'{\i}sica,
             38206 La Laguna, Tenerife, Spain \\     
         \and
               Research School of Astronomy and Astrophysics,
                The Australian National University, Weston, ACT 2611, Australia    \\       
         \and
              Department of Physics and JINA Center for the Evolution of the Elements, University of Notre Dame, Notre Dame, IN 46556, USA \\
         \and
             Institut Utinam, CNRS UMR6213, Universit\'e de Franche-Comt\'e,
        Observatoire de Besan\c con, Besan\c con, France \\ 
         \and
             Department of Astronomy and Astrophysics, The Pennsylvania State University, University Park, PA 16802, USA \\            
          \and
             Institute for Gravitation and the Cosmos, The Pennsylvania State University, University Park, PA 16802  \\
         \and
             Department of Astronomy and Space Science, Chungnam National University, Daejeon 305-764, Republic of Korea   \\         
         \and
             Apache Point Observatory, P.O. Box 59, Sunspot, NM 88349-0059, USA \\
             }

   \date{Received December 2014; accepted xxxx}

 
  \abstract
   {}
   {We analyze a sample of 3,944 low-resolution ($R \sim 2000$) optical spectra from the Sloan Digital Sky Survey (SDSS), focusing on stars with effective temperatures
 5800 $ \le T_{\rm eff} \le $ 6300~K,
and distances from the Milky Way plane in excess of 5 kpc,
and determine their abundances of Fe, Ca, and Mg.} 
   {We followed the same methodology as in the previous paper in this series,
 deriving atmospheric parameters by $\chi^2$ minimization, but this time we obtained
the abundances of individual elements by fitting their associated spectral lines. 
Distances were calculated from absolute magnitudes obtained by a statistical
comparison of our stellar parameters with stellar-evolution models.}
   {The observations reveal a decrease in the abundances of iron, calcium, and magnesium at large distances from the Galactic center. The median abundances for the halo stars analyzed are fairly constant up to a Galactocentric distance $r \sim$ 20 kpc, rapidly
decrease between $r \sim$ 20 and $r \sim$ 40 kpc, and flatten out to significantly lower values
at larger distances, consistent with previous studies. In addition, we examine the [Ca/Fe] and [Mg/Fe] as a function of [Fe/H] and Galactocentric distance. Our results show that the most distant parts of the halo show a steeper variation of the [Ca/Fe] and [Mg/Fe] with iron. We found that at the range $-1.6 <$ [Fe/H] $< -0.4$ [Ca/Fe] decreases with distance, in agreement with earlier results based on local stars. However, the opposite trend is apparent for [Mg/Fe]. Our conclusion that the outer regions of the halo are more
metal-poor than the inner regions, based on in situ observations of distant stars, agrees with recent results based on inferences from the kinematics of more local stars, and with predictions of recent galaxy formation simulations for galaxies similar to the Milky Way.
}
{}
   
   \titlerunning
   \authorrunning


   \maketitle
%

\section{Introduction}

In the half century since metal-poor stars were identified in the
halo of the Galaxy (Chamberlain \& Aller 1951), a great deal of effort
has been devoted to clarifying their origin, detailed chemical
abundance patterns, and observed kinematics (Beers \& Christlieb 2005;
Ivezi{\'c} et al. 2012; Frebel \& Norris 2013). One long standing debate
was focused on whether the Galaxy was formed from the monolithic
collapse of a protogalactic cloud (Eggen et al. 1962) or assembled from
smaller substructures (Press \& Schechter 1974; Searle \& Zinn 1978).
The current $\Lambda$CDM scenario predicts that large galaxies, such as
the Milky Way, formed hierarchically. If the Galactic halo formed from
accreted subsystems, at least some spatial and kinematical substructures
are expected to be observable (Helmi 2008; Klement 2010). This appears
to be the case for the outer regions of the Galaxy, but the same
scenario may not apply to the inner regions (Bell et al. 2008, 2010;
Schlaufman et al. 2009, 2011, 2012, and references therein; Xue et al.
2011), which are expected to be more completely phase-mixed.

Chemical tagging of the Galaxy's building blocks (Freeman \&
Bland-Hawthorn 2002) could provide important information about the
evolution of the different components of the Milky Way. In particular,
differences in the observed [$\alpha$-element/Fe]  for
halo stars could be explained in terms of the different star-formation
histories of the progenitor subgalactic systems that contributed
the stars found in the halo today. Type-II SNe produce
$\alpha$-elements and iron, and explode on short timescales ($\sim
10^{7}$ yr). Type-Ia SNe are thought to be the primary contributors
to the iron abundance in the interstellar medium, but their
evolutionary time scales are much longer ($\sim10^{9}$ yr).

During the past decade a number of authors have reported the possible
existence of a dichotomy in the abundances of halo stars in the solar
neighborhood, often related with their observed kinematics. Fulbright
(2002), for example, found a correlation between the observed space
velocities of a relatively small sample of stars (with metallicities in
the range $-2.0 \le $ [Fe/H] $ < -1.0$) and their [$\alpha$/Fe];
stars with lower [$\alpha$/Fe] were argued to be associated with faster space motions.
Gratton et al. (2003) reported that the stars in their sample with
substantial prograde rotation about the Galactic center exhibited higher
[$\alpha$/Fe] than those with small or retrograde rotation, as
confirmed later by Jonsell et al. (2005). Ishigaki et al. (2010) found
that in the metallicity range $-2 <$ [Fe/H] $< -1$, stars on orbits
reaching a maximum distance from the Galactic plane $|Z| > 5$ kpc
possess [Mg/Fe] $\sim$0.1~dex lower than those that only reach $|Z| < 5$
kpc. 

The results of Nissen \& Schuster (2010, 2011) reinforced the
claims that there are (at least) two distinct stellar populations in the halo. Based
on high-resolution spectroscopy of a very local sample of moderately
low-metallicity stars, $-1.6 <$ [Fe/H] $< -0.4$, these authors observed
two different trends in the [$\alpha$/Fe] with metallicity, one
higher and flatter, and the other comprising stars with lower mean
[$\alpha$/Fe] and a steeper slope, associated with stars having a
higher velocity dispersion. Their suggested interpretation calls for an
inner, old, flattened high $\alpha$-element population with a prograde
rotation, formed during a phase of dissipative collapse, and an outer,
slightly younger spherical population, exhibiting counter-rotation, with
lower values of [$\alpha$/Fe], and presumably accreted from (relatively
massive) dwarf-like galaxies.  

These previous samples included relatively small numbers of stars,
exploring a limited range of metallicity and kinematic phase-space. 
Recent spectroscopic surveys have been able to provide more
comprehensive results based on their dramatically larger samples of
stars. For example, the Sloan Digital Sky Survey (SDSS: York et al.
2000), in operation since 2000, and its extensions (SDSS-II: Abazajian
et al. 2009; SDSS-III: Eisenstein et al. 2011), includes the
stellar-specific program SEGUE-1 (Sloan Extension for Galactic Understanding
and Exploration; Yanny et al. 2009), which was later extended with SEGUE-2. The total number of low-resolution stellar spectra
gathered by SDSS and SEGUE is on the order of 750,000. The stars used for flux calibration are of particular interest
because they were observed for every plug-plate and thereby benefit from the dense tiling of the SDSS
footprint.
 
Carollo et al. (2007,2010; see also Beers et al. 2012) made use of the
calibration stars available at the time to separate the halo into (at
least) two stellar populations, with clearly different spatial-density
profiles, stellar orbits, and metallicity distribution functions, which
they referred to as the inner-halo and outer-halo populations. According
to their interpretation, the inner-halo population formed as the result
of the dissipational mergers of (relatively more massive) subgalactic
fragments, while the outer-halo population formed from dissipationless
mergers of (relatively less massive) fragments. Their results have found
observational support from analyses of photometric samples of SDSS stars
(which are not subject to possible target-selection biases in the
spectroscopic samples) in the work of de Jong et al. (2010) and An et
al. (2013).

Of key importance in the analysis of Carollo et al. is the derivation of
kinematical properties for the entire halo from a relatively local sample of calibration
stars ($d \le 4$ kpc), for which reasonably accurate proper motions
could be obtained. The distance estimates that they employed and used
in concert with the observed radial velocities and proper motions to
derive space motions have been criticized by Sch{\"o}nrich et al. (2011, 2014), who called into
question their inferred division of the halo into inner and outer components. Beers et
al. (2012) rejected this claim and presented additional evidence in
support of the case for the dual halo.

To resolve this situation, and independent evaluation based on an in situ halo sample can shed light on the matter. In the present series of papers we conduct such an analysis, based on
the original SDSS/SEGUE supplemented by new
spectra obtained as in the course of the Baryon Oscillations Spectroscopic
Survey (BOSS: Dawson et al. 2013), which is part of SDSS-III. In Paper I
of this series (Allende Prieto et al. 2014), we analyzed a sample of the spectrophotometric
calibration stars, including targets at large distances (in excess of 10
kpc) at high Galactic latitudes observed in BOSS. Stellar
parameters and metallicity were derived, providing a first look at the
in situ metallicity distribution function of halo stars from the deepest
analysis yet obtained. Here we extend this sample and derive
not only [Fe/H] from iron lines, but also abundances for two $\alpha$-elements, [Ca/H] and
[Mg/H]. In Sect. 2 we describe our sample. Sect. 3 explains our
methodology and verifies the accuracy of the derived parameters. Our
results are summarized in Sect. 4, followed by a discussion and
conclusions in Sect. 5.

\section{Observations}
\label{observations}

We employed stellar spectra from the tenth SDSS data release (DR10; Ahn et
al. 2014), which contains SEGUE-1 and SEGUE-2 data, calibration stars from
earlier SDSS observations, and those taken over the last four years as
part of BOSS. The spectra were obtained with the 2.5m telescope at
Apache Point Observatory (Gunn et al. 2006), using a pair of double
spectrographs connected to 640 (SDSS, SEGUE-1, SEGUE-2) or
1000 optical fibers (BOSS; see Smee et al. 2013), as explained in more
detail in Paper I.

SEGUE-1 was conceived to explore the different stellar populations of
the Galaxy, in particular to study the chemistry and kinematics of the
spatial substructure found in the stellar halo from photometry in the
original SDSS program. Stars at large distances (tens of kpc) were
observed over a broad range of apparent magnitude (14.0 $< g <$ 20.3)
and Galactic latitude in order to extensively cover the large
structures. A wide range of spectral types were included, from F/G, G,
dK, dM, and type-L brown dwarfs, including a substantial number of
thick-disk stars in the solar neighborhood. SEGUE-2 concentrated
primarily on halo stars, increasing the number of red giant-branch (RGB)
stars and blue horizontal-branch (BHB) stars, reaching distances
of up to
100 kpc from the Sun. The spectra have low resolution, $R \equiv
\lambda$/FWHM($\lambda$) vary over $1500 < R < 2500 $, and cover 
the wavelength range 3900\,{\AA} $< \lambda < $ 9000\,{\AA}.

The BOSS spectrophotometric calibration stars are mainly halo
main-sequence turnoff (MSTO) stars at high Galactic latitudes and reach
large distances into the halo, up to 100 kpc in some cases. They
were obtained using an upgraded version of the spectrographs that
resulted in increased sensitivity, an enlarged wavelength coverage
(3600\,{\AA} $< \lambda <$ 10000\,{\AA}), and a resolution similar to the original spectrographs.

\section{Analysis}
\label{analysis}

The goal of this work is to investigate the observed distribution of the
abundances of several elements in the Galaxy halo, in particular
Fe, Ca, and Mg. Our strategy, following the work described in Paper
I, is to determine these abundances from stars observed in situ and
to examine how they vary as a function of distance. In addition, we study
the [Ca/Fe] and [Mg/Fe] as a function of metallicity and
distance from the center of the Galaxy, $r$, to examine whether
different trends are detected throughout the halo.

\subsection{Determining abundances.}
\label{det_ab}

We determined the abundance of an element following the techniques
explained in Paper I, which employ an updated version of the code FERRE\footnote{FERRE is available from http://hebe.as.utexas.edu/ferre} (Allende Prieto et al. 2006). This code searches for the
model spectrum in an n-dimensional grid of synthetic spectra that best
fits the observed spectrum, using $\chi^{2}$ as the merit function and
interpolating with a Bezier scheme. The search returns the
values of the stellar atmospheric parameters of this synthetic spectrum.
It is possible to fix some of these values before starting, so that the
search is limited to the parameters of interest.

The analysis performed in Paper I used the entire spectrum in the fit to
determine the atmospheric parameters (effective temperature $T_{\rm eff}$, surface gravity in logarithmic units $\log g$, and metallicity [M/H]\footnote{[M/H]=$\log_{10}$(M/H)-$\log_{10}$(M/H)$_{\rm Sun}$}). In the case of metallicity,
the fit of the entire spectrum implies that several elements are involved in its determination. For example, at low
metallicities, the Ca II resonance doublet and the Mg I$b$ triplet at
5180\,{\AA} are the main contributors to the metallicity determination,
since they are the strongest features. However, if we only fit iron
lines, we achieve a more direct, and provided they are measurable, more reliable
determination of the iron abundance. Moreover, as the abundances of all
the elements scale with the iron abundance in the generation of the
synthetic spectra (for the $\alpha$-elements an increase at low
metallicities is taken into account -- see Paper I), searching for
metallicity by fitting lines that belong to other elements permits determining their abundances. In this manner, we determined
estimates of the Fe, Ca and Mg abundances. Note that a different
relation between the abundances of the elements in the composition of a
star would affect the opacities and the equation of state, making this
approximation fail. However, this does not seem the case for most of our
program stars, and will not affect our statistical results.

We performed an analysis of our sample (BOSS spectra and all previous
optical SDSS spectra that satisfied the condition of having a redshift
$|z| \le 10^{-2}$, 55,401 from BOSS and 645,354 from SDSS/SEGUE), using a grid of synthetic spectra covering the
following stellar parameters and metallicity ranges: 4750~K $< T_{\rm
eff} < 6500$~K, $0.5 < \log g < 4.5$ and $-5.0 <$ [Fe/H] $< +0.5$. This
is the same grid as described in Paper I. We selected objects that
were assigned an effective temperature between
5800 and 6300 K and $\chi ^{2} < 10$ in this first analysis. Because of the systematic errors identified in Paper I for low gravity stars, we only considered stars with $\log$ g $>$ 2.5. This includes 50,252 from BOSS and 95,536 from SDSS/SEGUE. To measure the chemical abundances of Fe and the two
$\alpha$-elements, Ca and Mg, we performed a second analysis in which we fit
features dominated by transitions of a single element at a time, fixing
$T_{\rm eff}$ and $\log g$ to the values obtained in the first analysis. 

\begin{figure*}
\centering
\includegraphics[width=4.5cm,angle=90]{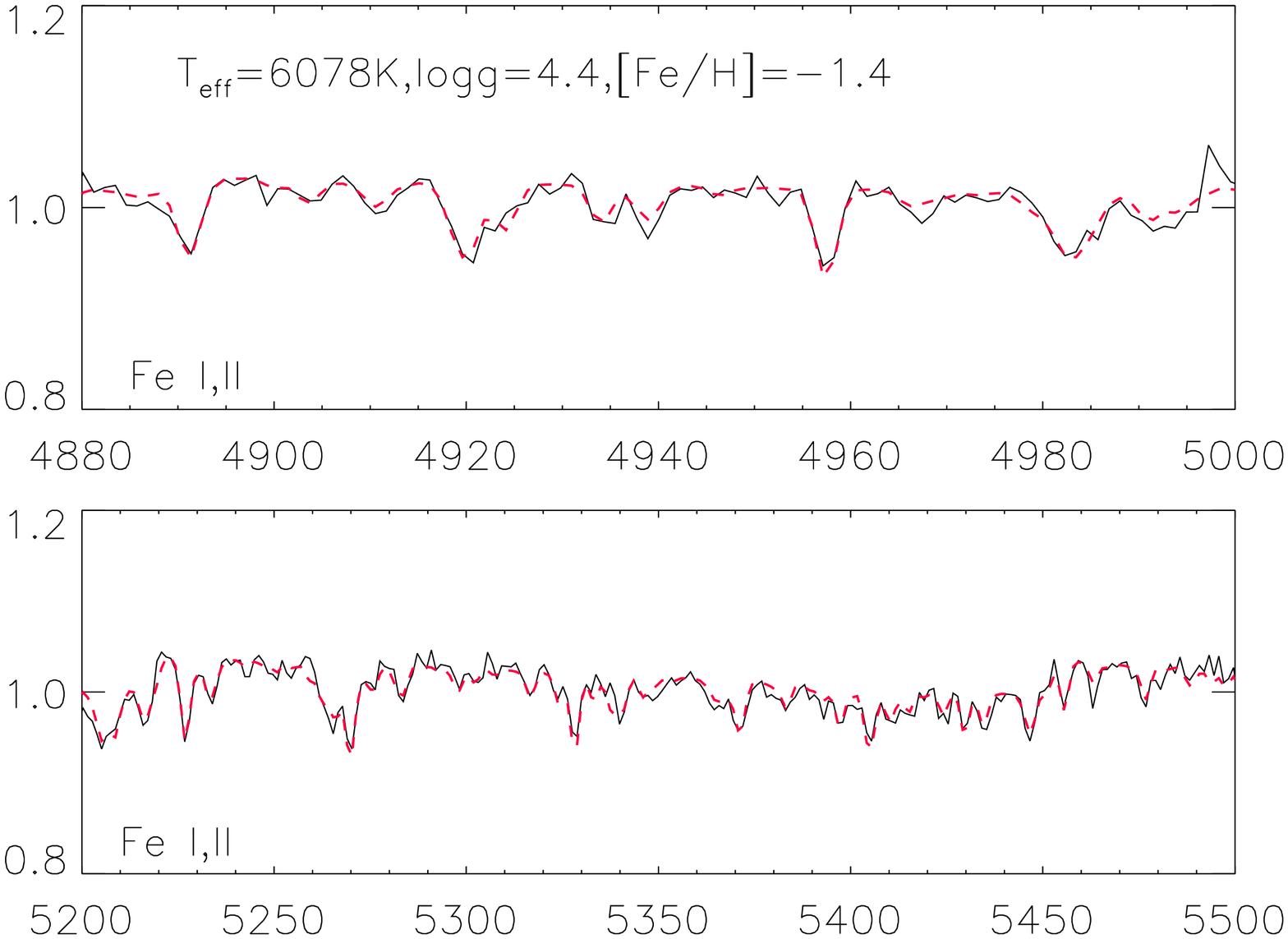}
\includegraphics[width=4.5cm,angle=90]{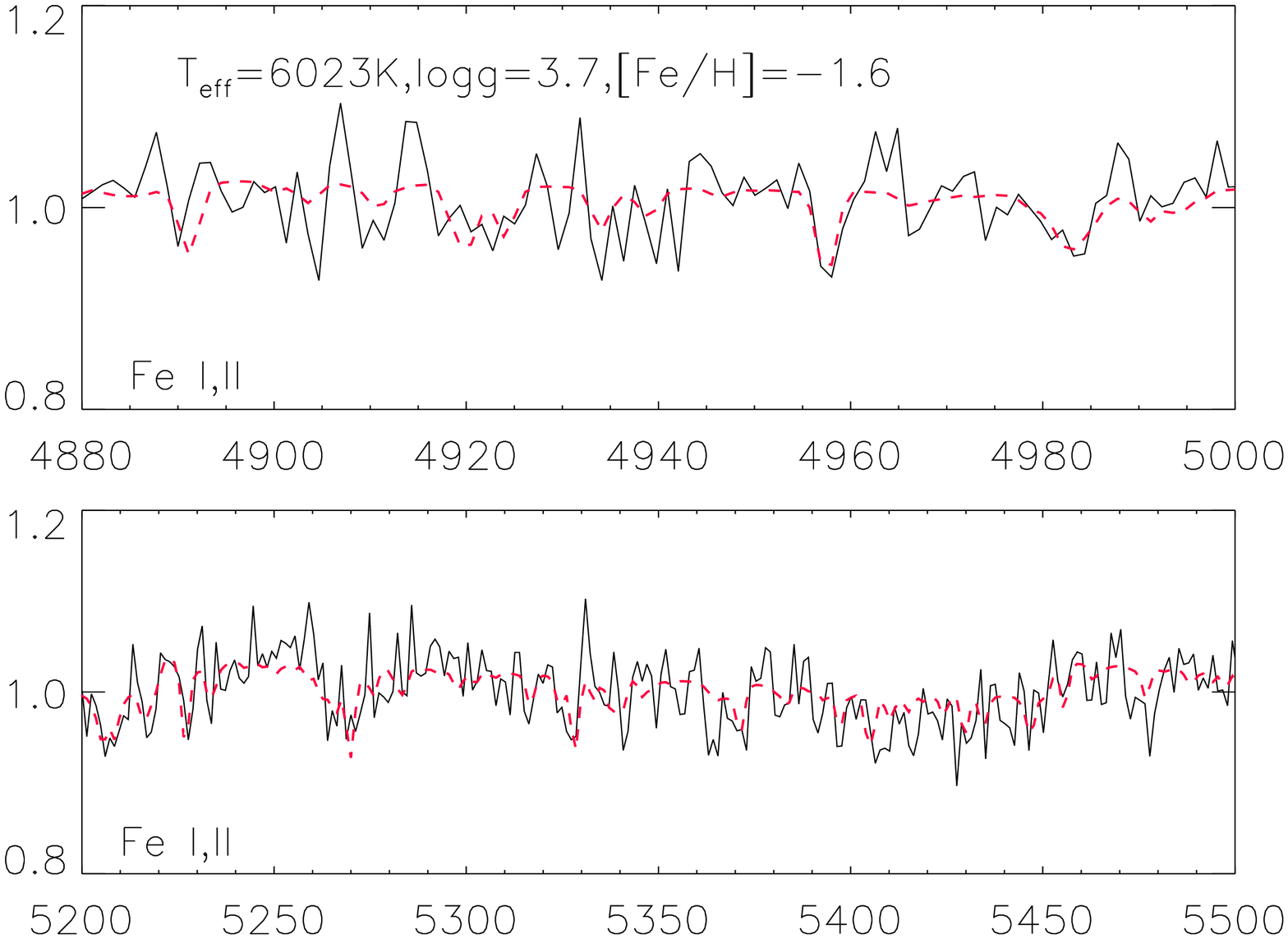}
\includegraphics[width=4.5cm,angle=90]{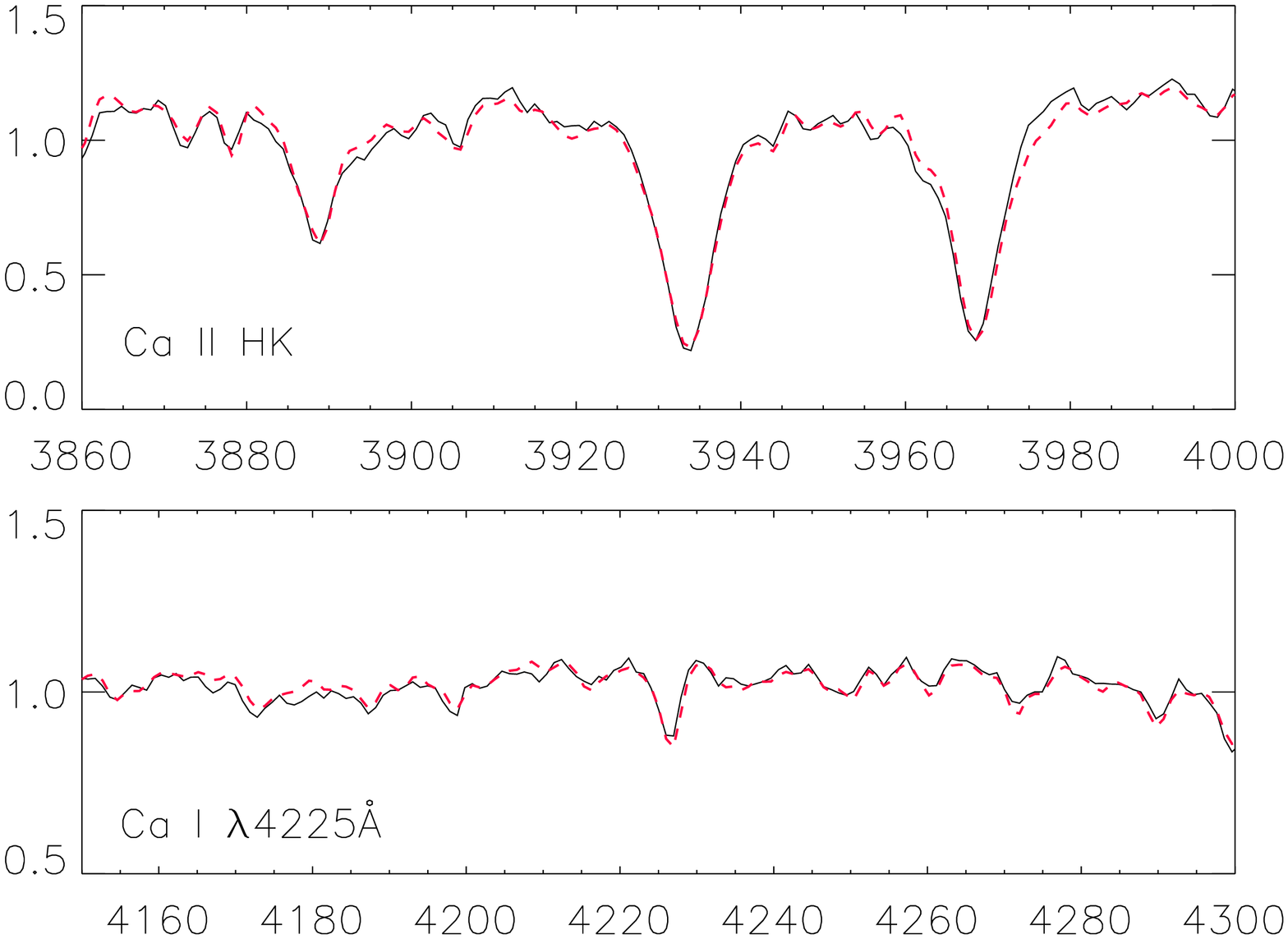}
\includegraphics[width=4.5cm,angle=90]{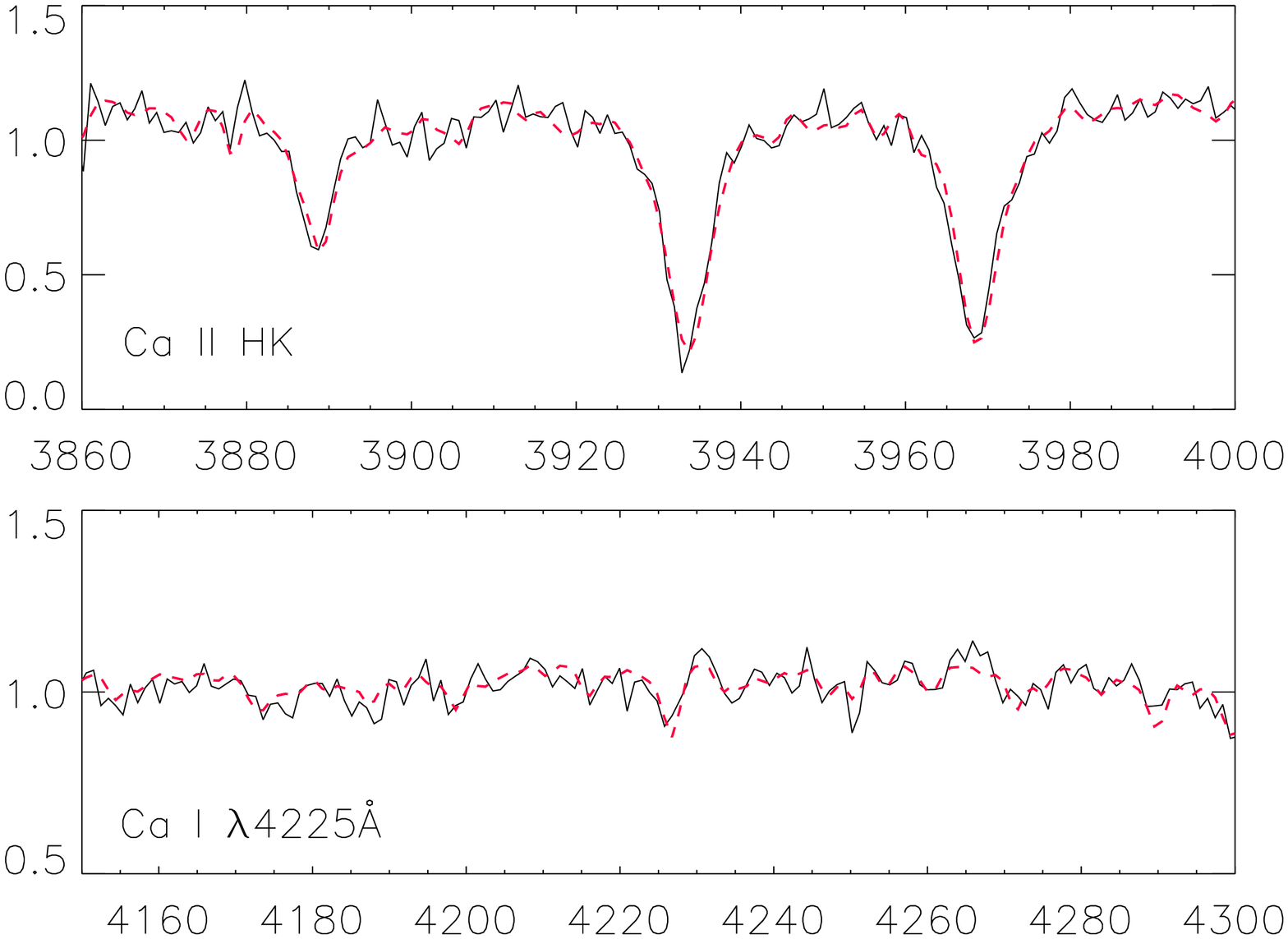}
\includegraphics[width=4.5cm,angle=90]{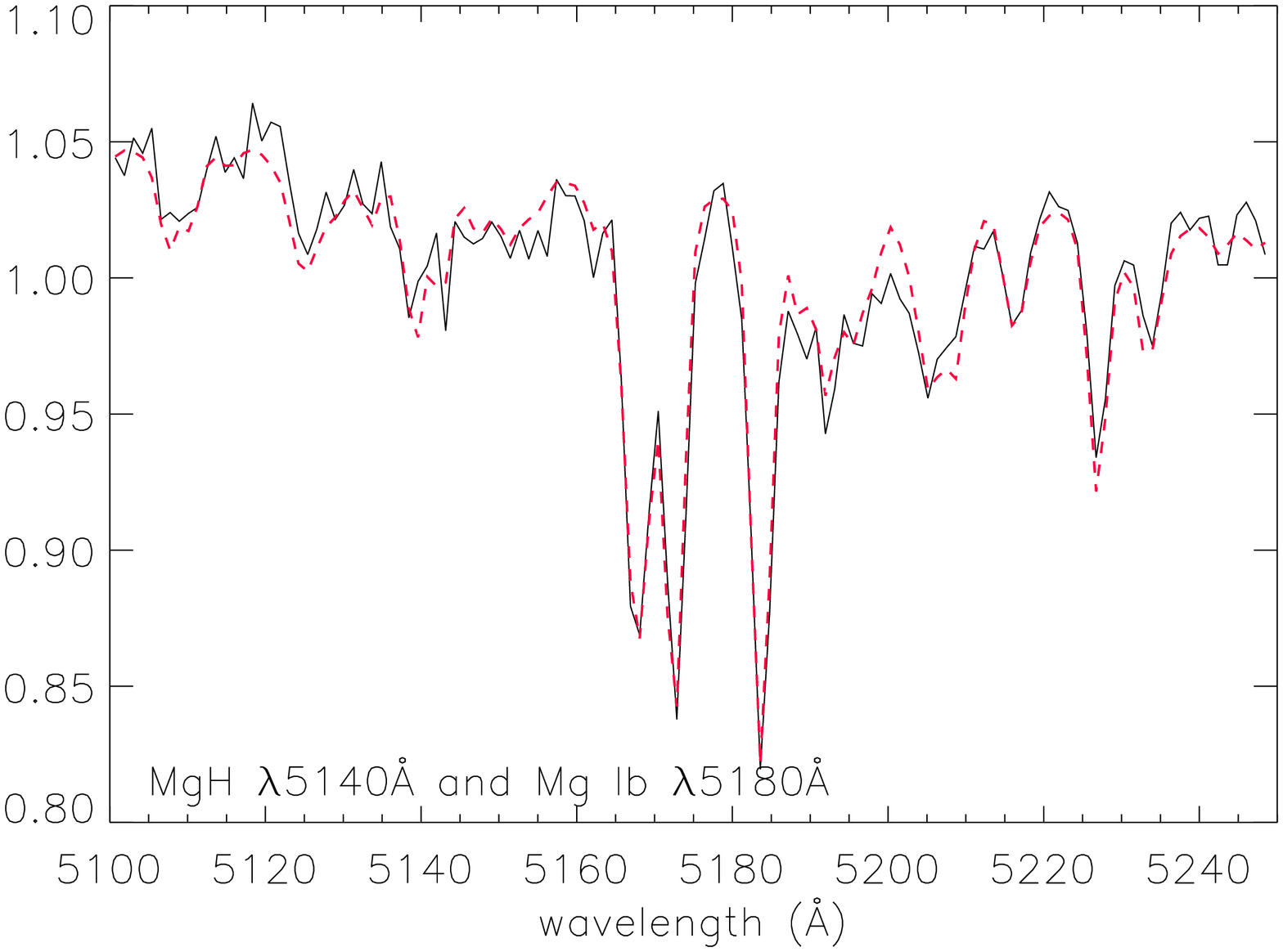}
\includegraphics[width=4.5cm,angle=90]{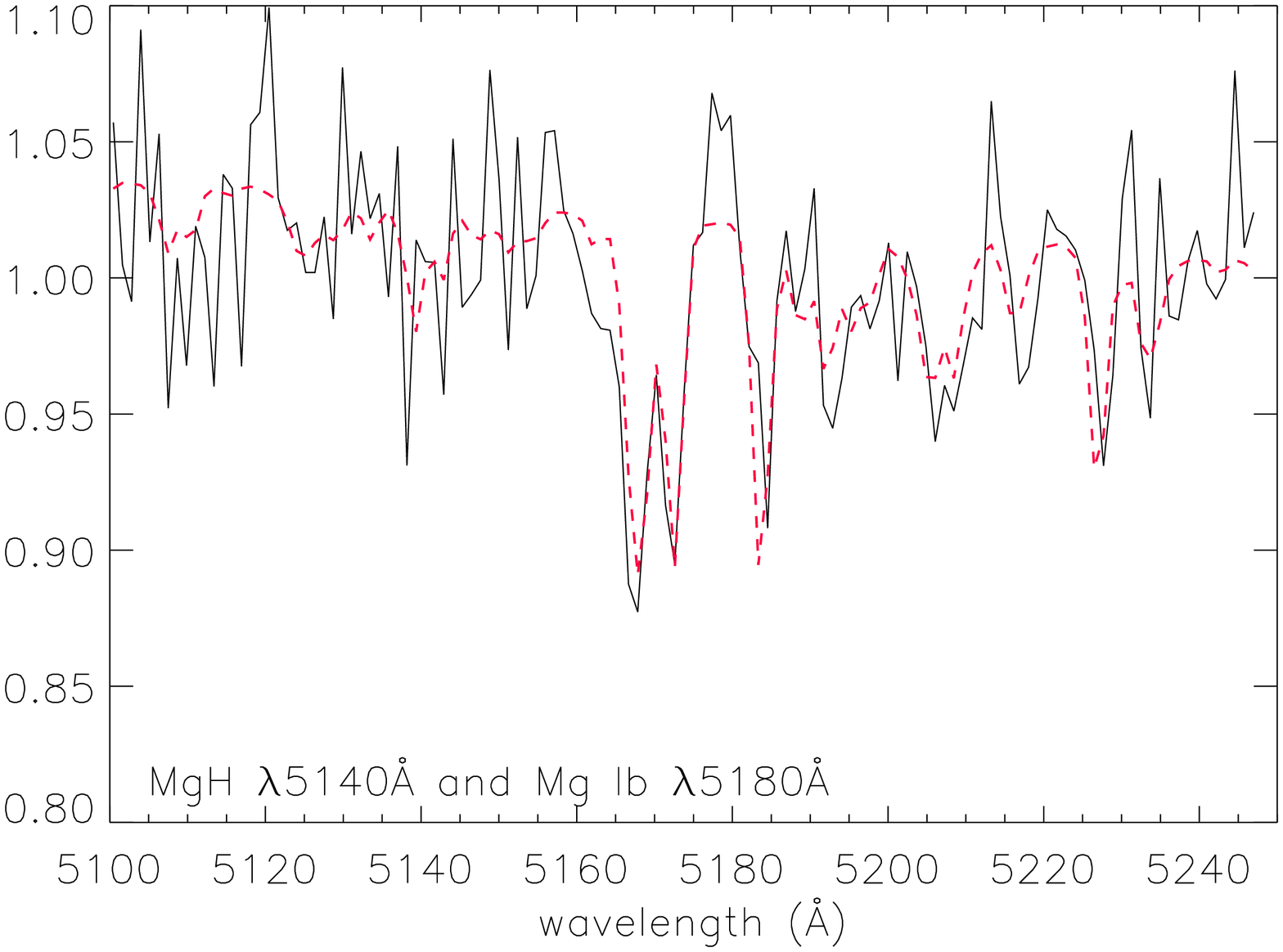}
\caption{Fit of the regions for two example spectra (with atmospheric 
parameters $T_{\rm eff}$ = 6078~K, $\log g$ = 4.4, and [Fe/H] $= -1.4$ and $T_{\rm eff}$ = 6023~K, $\log g$ = 3.7, and [Fe/H] $= -1.6$)
that contains some of the Fe, Ca, and Mg lines that contribute most to
the determination of their abundances. The red line corresponds to the
fit spectrum, plotted over the observed spectrum (in black). The panels on the left show an spectrum with a S/N $\sim 60$; the panels on the right correspond to a spectrum with the median S/N of the data, $\sim 30$.}
\label{ajustefe}
\end{figure*}

We chose the most suitable Fe, Ca, and Mg lines to derive the corresponding abundances by identifying which features are most sensitive to abundance changes and were measurable at low spectral resolution. The
sensitivity was determined by calculating the ratio between two
synthetic spectra, generated for a given set of values for the effective
temperature, gravity, and metallicity, but with a difference of 0.1 dex
in the abundance of Fe, Ca, or Mg. For this test, we adopted $T_{\rm
eff}$ = 6000 K, $\log g$ = 4.0, and [Fe/H] $= -1.0$. The ratio indicates
the features with the highest sensitivity to a given abundance change.
Figure~\ref{ajustefe} shows some of the most relevant lines used to
derive the abundance of each element. Our analysis used the same model grid as in Paper I, but we selected only the specified regions that are most sensitive to
the individual abundances.

\begin{figure*}
\center
\includegraphics[angle=90,scale=0.6]{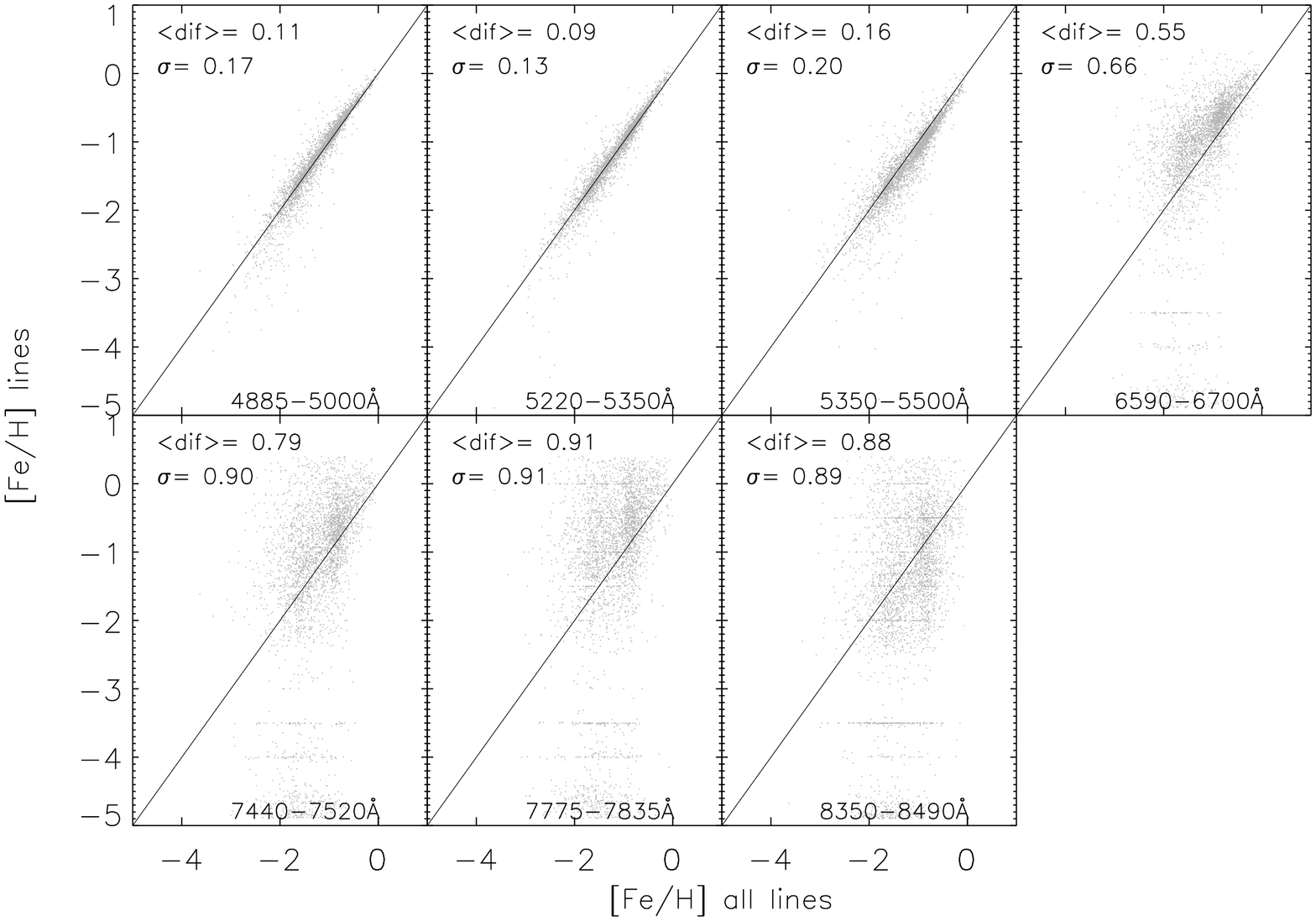}
\caption{Correlation between the values of [Fe/H] obtained by fitting
different spectral regions containing Fe lines, compared with the
estimation obtained by fitting all of
the selected lines together. The bluest regions best contribute to the
determination of [Fe/H], with rms values lower than 0.2 dex. The results
of the fitting line for the four reddest regions exhibit little
correlation with estimates using all of the regions simultaneously.}
\label{diffe}
\end{figure*}

\begin{figure*}
\center
\includegraphics[angle=90,scale=0.6]{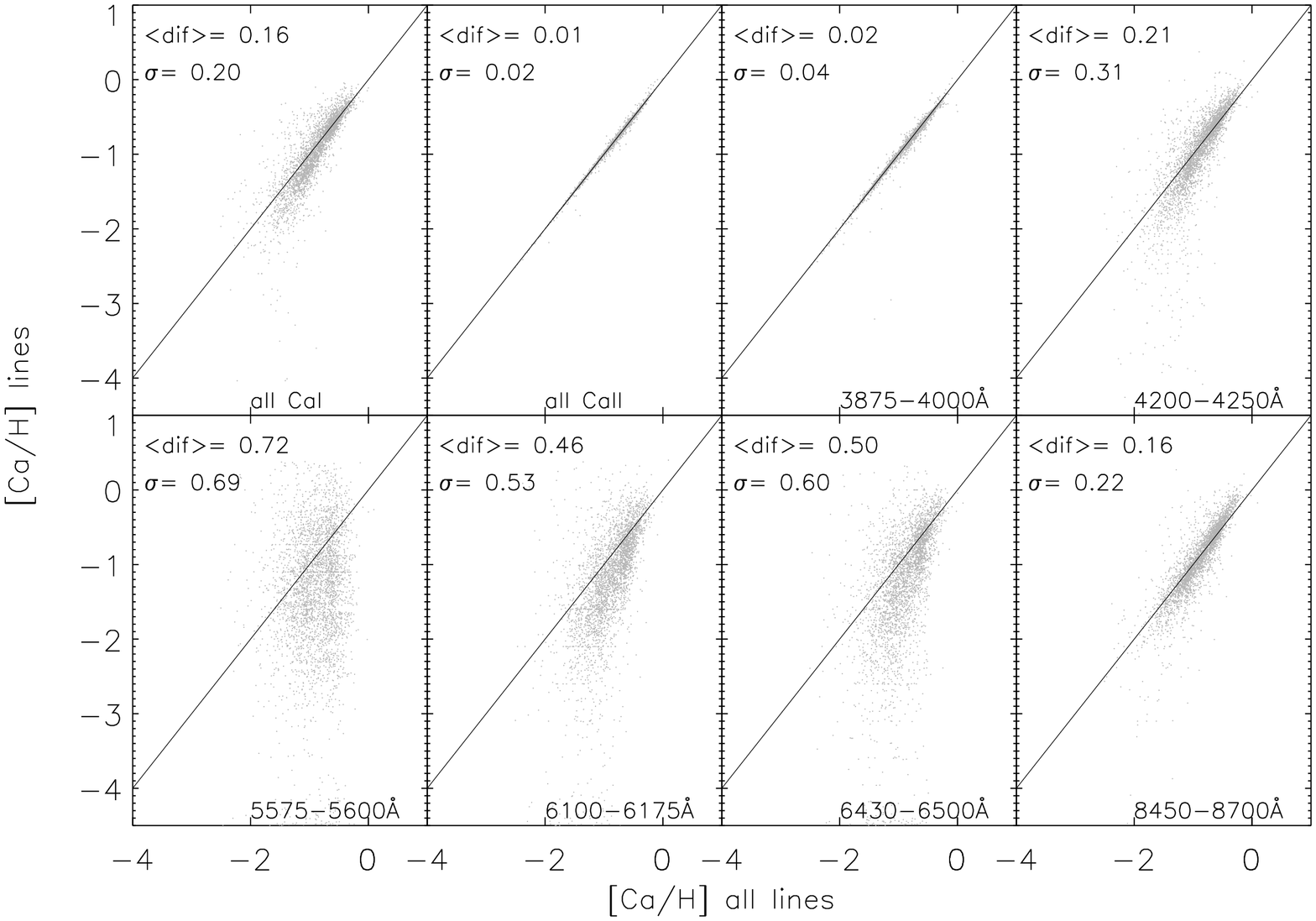}
\caption{Correlation of the [Ca/H] results calculated for a sample of 61,591 stars, using different spectral regions
containing Ca lines (neutral and/or ionized), compared with the estimation
obtained by fitting all of the selected lines together. The CaII lines are
the most sensitive to changes in the abundance of calcium, in particular the Ca HK doublet, hence they are the
primary features used. Of the neutral lines,
the CaI line at 4226.73\,{\AA} is the most relevant.}
\label{difca}
\end{figure*}

In the case of Fe and Ca, several regions are sufficiently sensitive for
use in deriving accurate abundance estimates. Although lines in the blue
portion of the spectrum tend to show higher sensitivity, some lines in the red
were chosen and analyzed as well. We performed several analyses, fitting
one of the selected regions in isolation, and another fitting all
the regions together. In the first case, the continuum was determined by dividing each region by its mean, and the fit was performed by calculating the $\chi^{2}$ using only a percentage of the region that includes the lines of interest and avoiding others that do not belong to the element. In the latter, these lines of interest were fit over the spectrum normalized as explained in Paper I, that is, by splitting the
spectra into 200\,{\AA} bins and dividing the fluxes in each bin by the
mean values. In Table~\ref{tbl-1} we list the selected spectral regions, the dominant lines in the regions, and the percentage of each
region that was fit. We only considered estimates for which
$\chi^{2} < 10$, as in the analysis of the full spectrum, and a signal-to-noise ratio higher than 20, considering the median value of this ratio in the available spectral range. The comparison of the results obtained for each
short region containing one or a few lines with the results obtained by
fitting all of the lines simultaneously throughout the spectrum simultaneously
(Figs.~\ref{diffe} and \ref{difca}) provides the means to estimate the
uncertainties in our determinations and a way to identify which regions in the spectrum
contribute the most to the derivation of the abundance of a given
element. For Mg, we performed only one analysis, fitting the
blend of the MgH band at about 5140\,{\AA} and the Mg I$b$ triplet at
$\sim$5165-5190\,{\AA}.

Figure~\ref{diffe} includes results for the SEGUE-1 and SEGUE-2 samples and earlier SDSS stellar spectra. This figure compares the iron abundance estimates from the fit of only one of the selected regions with the results obtained by fitting all the regions together for a sample of 46,065 objects. Inspection of
this figure indicates that the Fe lines that contribute the most to
determining [Fe/H] are those in the range of 5220-5350\,{\AA}, with
a standard deviation $\sigma \sim$ 0.13 dex. Two other windows exhibit a
$\sigma$ lower than 0.20 dex, but the results for the four reddest
regions show little correlation with the [Fe/H] values obtained using all the windows together.
However, we have verified by comparing with the [M/H] values obtained in Paper I that the estimates including all of the
regions in the fitting are slightly more precise than those considering only one
or a few of them, obtaining an offset of 0.11 dex and $\sigma = 0.14$ dex, whereas 
an analysis excluding the four reddest regions shows an offset of 0.13 dex and $\sigma = 0.14$ dex. For Ca (see Fig.~\ref{difca}) the CaII lines
are the main contributors in determining [Ca/H], with rms
deviations lower than 0.05 dex compared with measurements from all the
calcium lines. Fitting only neutral lines results in an rms deviation of
0.20 dex, which increases at low metallicity. The most relevant feature
of the neutral lines is the CaI line at 4226.73\,{\AA}.

\begin{figure*}
\center
\includegraphics[width=6.5cm]{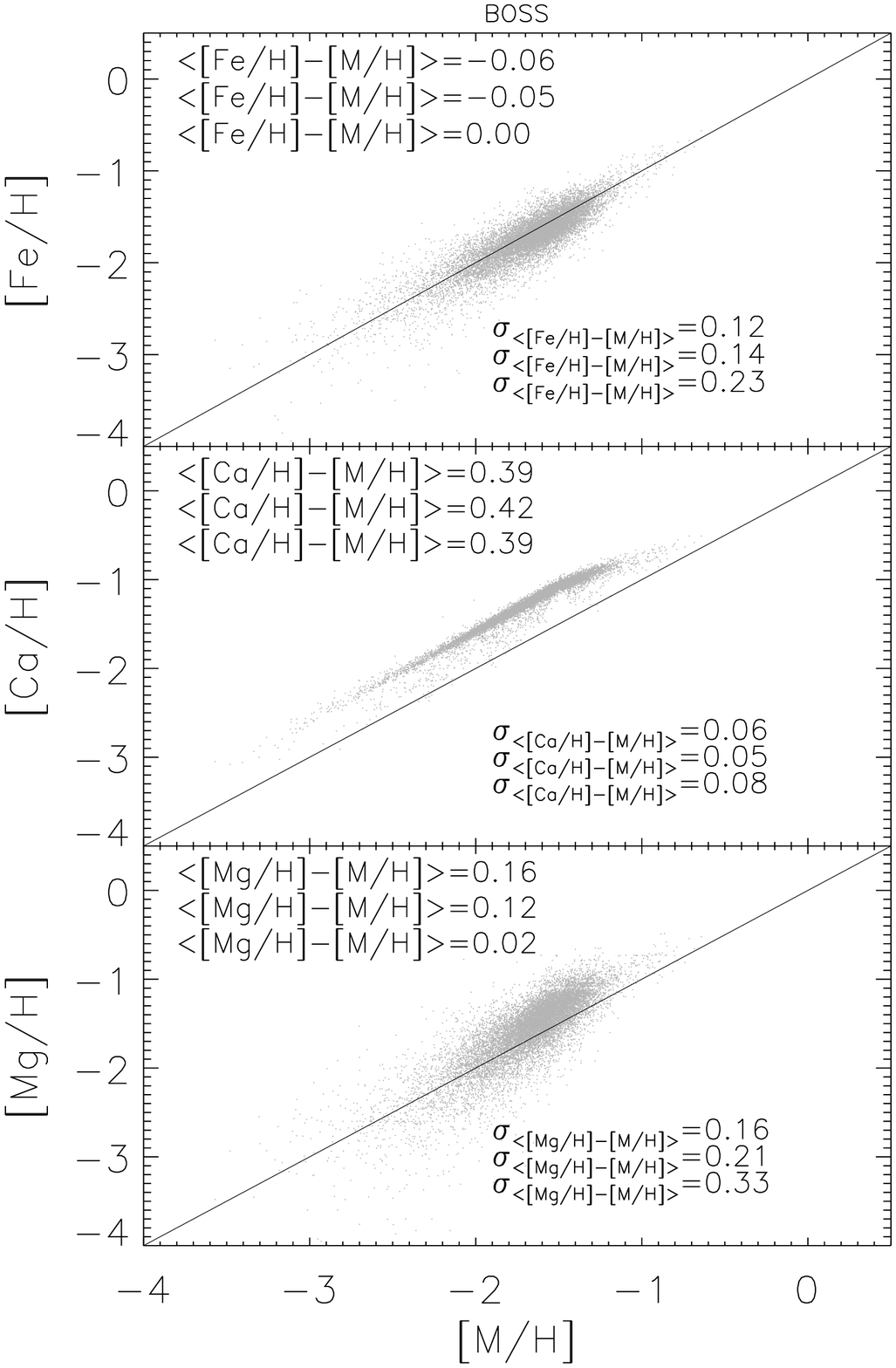}
\includegraphics[width=6.5cm]{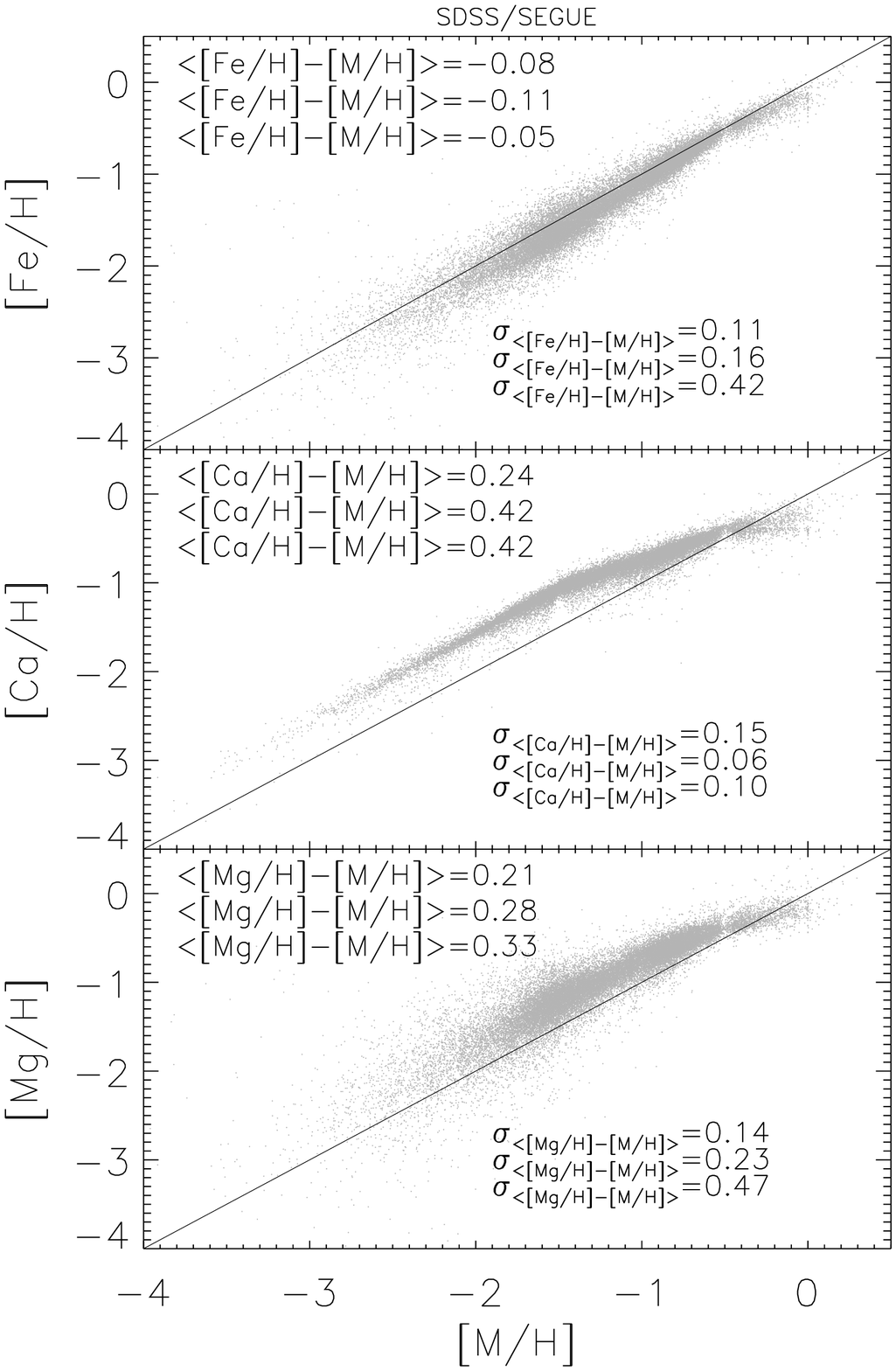}
\caption{Comparison of the [Fe/H], [Ca/H], and [Mg/H] estimates from
SDSS/SEGUE and BOSS stars obtained by fitting individual lines with the
metallicity results determined in Paper I, [M/H], which considered the entire
spectrum in the analysis. The mean difference between x and y axes appears in the top left corner of each panel, evaluated in three ranges of [M/H]: [M/H] $> -1.5$, $-2 <$ [M/H] $< -1.5$, and [M/H] $< -2$, and the corresponding standard deviations are written in the bottom right corner. The top panels show good agreement
between [Fe/H] inferred from individual lines of this element, with an offset of
$\sim$ 0.1 dex. [Ca/H] estimates are offset by $\sim0.4$ dex over [M/H], except at metallicities
higher than [M/H] $= -1.5$, as expected. [Mg/H] estimates display a
higher dispersion, increasing with decreasing metallicity, and lower deviations for BOSS stars with respect to SDSS/SEGUE stars. 
}
\label{comp_MH_boss}
\end{figure*}

We now examine correlations of [Fe/H], [Ca/H] and [Mg/H] with the global
metallicity, [M/H], derived in Paper I, which was calculated by fitting the entire
spectrum. The top panels in Fig.~\ref{comp_MH_boss} reveal
an offset of about 0.1 dex between the overall metallicity [M/H] and the
iron abundances inferred exclusively from iron lines, with a dispersion
that increases at low metallicities. The values of [Ca/H] 
are offset by about $0.4$ dex over [M/H], except at metallicities higher than
[M/H] $= -1.5$, as expected for metal-poor stars. Conversely, the [Mg/H]
estimates exhibit a higher dispersion, increasing as the metallicity
decreases. The BOSS and SDSS/SEGUE stars exhibit similar behaviors.
However, the dispersion in [Fe/H] for the BOSS stars is lower for [M/H] $< -2.0$,
and the [Mg/H] determinations in BOSS spectra exhibit deviations with
respect to [M/H] that are significantly lower than for SDSS/SEGUE stars,
approaching zero for stars with [M/H] $< -2.0$.

\subsection{Determining distances.}
\label{detdist_sub}

To explore possible variations of the chemical trends with
distance, we estimated the distances for each star from the Galactic plane and from the 
Galactic center. To obtain the distance from the Sun, $d$, we
first derived the absolute magnitude, $M_{V}$, following Allende Prieto
et al. (2006). In this approach, the probability of a
given star to have the adopted values of T$_{\rm eff}$, $\log g$, and [Fe/H] is calculated by
comparing it with a set of isochrones (e.g., Bertelli et al. 1994). Using
this value of M$_{V}$, we calculated the distance from the equation

\begin{equation}
M_V = 5 + m_v - 5 \log d,
\end{equation}

\noindent where m$_{v}$ is the dereddened apparent magnitude $V$, 
determined from the relationship (Zhao \& Newberg 2006)

\begin{equation}
V = g - 0.561(g-r).
\end{equation}

\noindent The Galactocentric distance, r, was calculated using

\begin{equation}
r = \sqrt{d^{2} + R_{\odot}^{2} - 2d R_{\odot} \cos{b} \cos{l}}
\end{equation}

\noindent and the distance from the plane, $Z$,

\begin{equation}
Z = d\sin{b},
\end{equation}

\noindent adopting $R_{\odot}= 8$ kpc (Bovy et al. 2009).

\begin{figure*}
\center

  \includegraphics[width=4.5cm, angle=90]{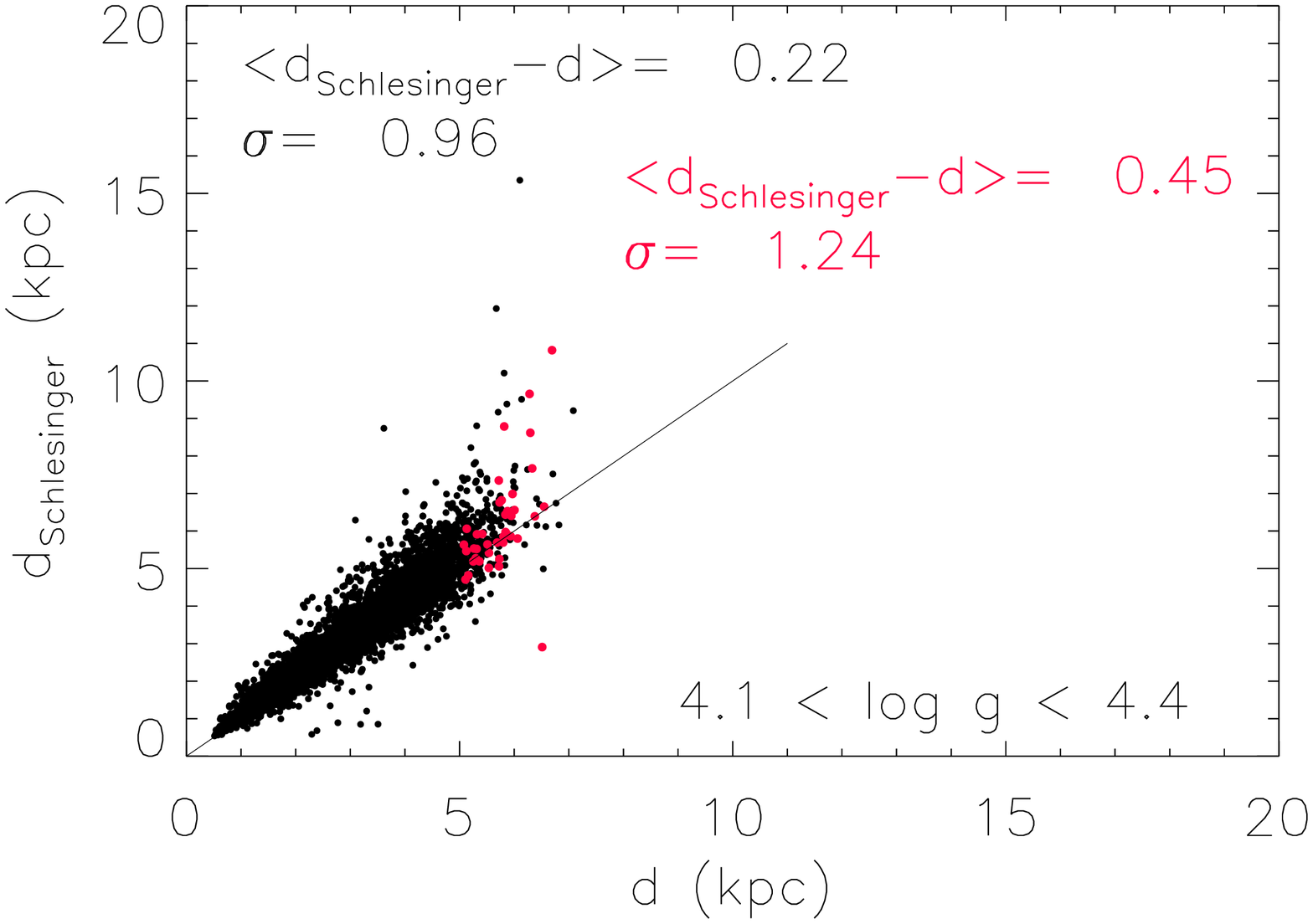}
  \includegraphics[width=4.5cm, angle=90]{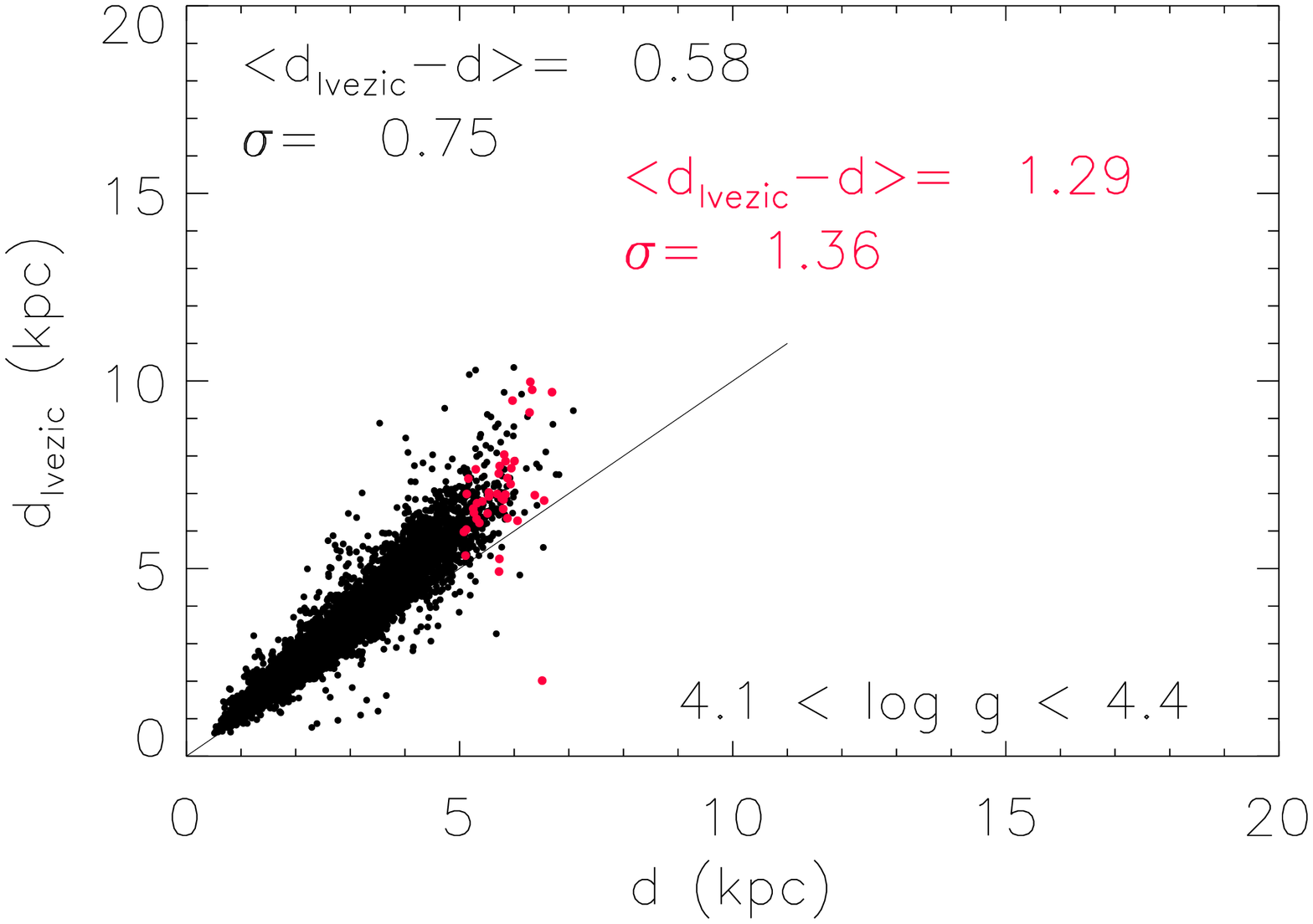}
  \includegraphics[width=4.5cm, angle=90]{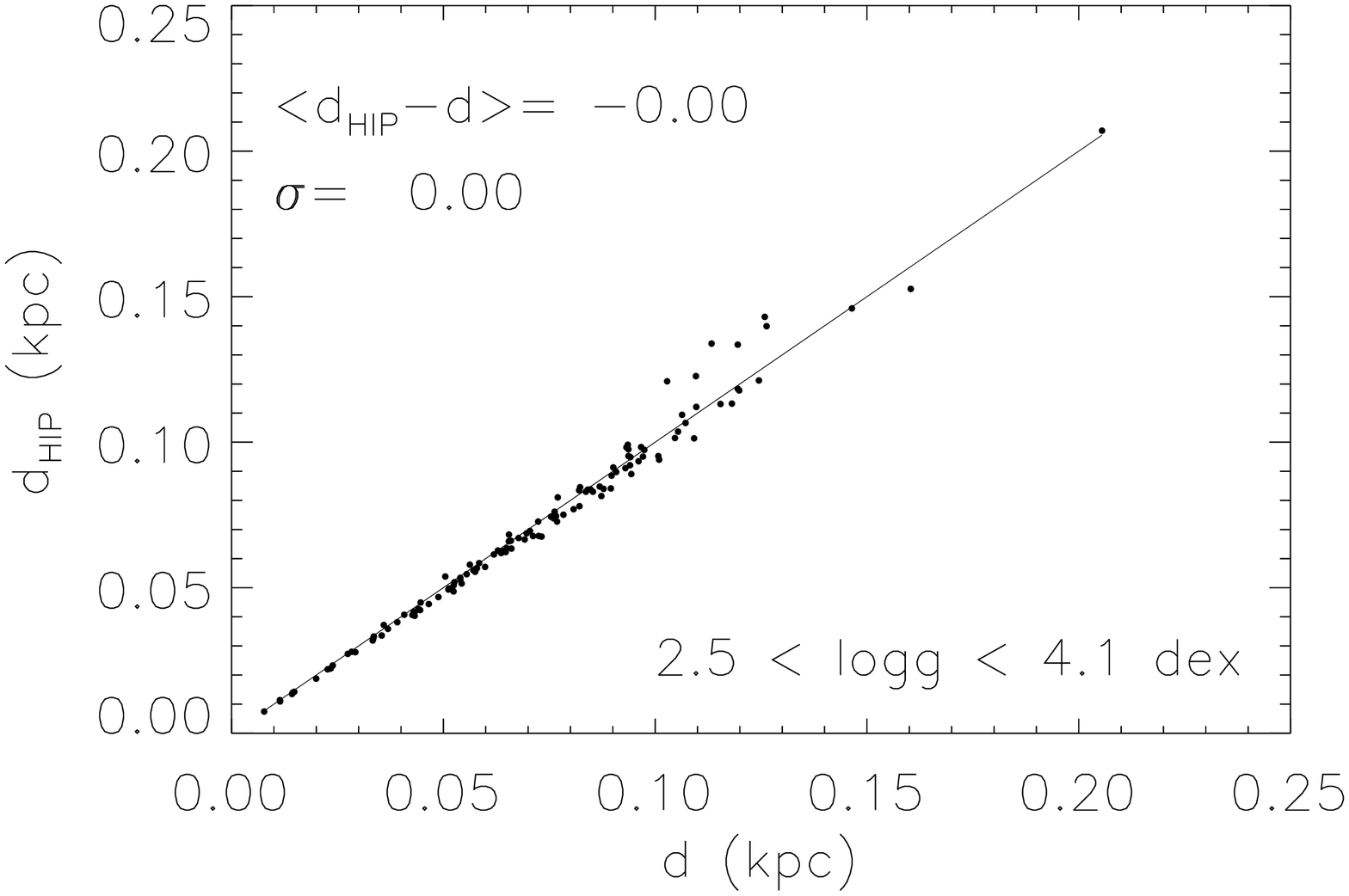}
  \includegraphics[width=4.5cm, angle=90]{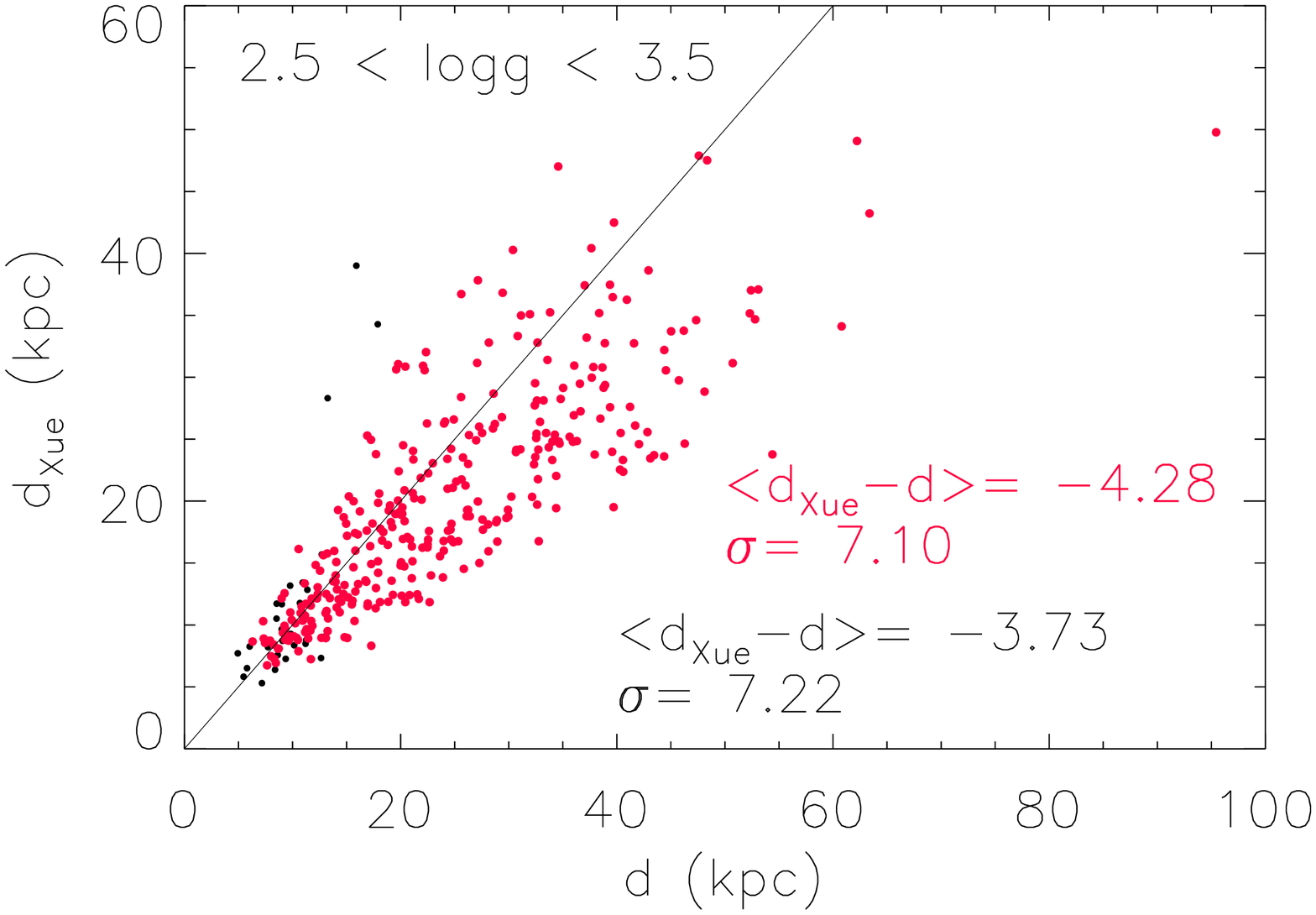}
  \caption{Comparison of distances to the Sun (for pre-BOSS stars), $d$,
calculated using the methodology followed in this work. The top-left panel
corresponds to the comparison of stars with $4.1 < \log g < 4.4$ and
$-2.0 < $ [Fe/H] $< -0.4$, with distances calculated using the approach
described by Schlesinger et al. (2012; see text). The top-right panel
shows distance estimates for the same sample following Ivezic et al.
(2008). In the bottom left panel, distances based on parallaxes from the
Hipparcos satellite are compared with our estimates for a sample of
turnoff and giant stars with stellar parameters and metallicities from
Ramirez et al. (2013) with $2.5 < \log g < 4.1$. The bottom right
panel shows the comparison of distances calculated for a sample of SEGUE
giants with $2.5 < \log g < 3.5$, compared with those obtained by Xue et
al. (2014). Objects at $|Z| > 5$ kpc, calculated from our distance estimations, as well as their statistics, are shown in red.}
\label{rdwarf_k}
\end{figure*}

We evaluated the reliability of our results using distances estimated
following other methods. A first validation was obtained by comparing
with 10 Gyr empirically-calibrated YREC isochrones (An et al. 2009).
Then, after identifying the closest isochrone in [Fe/H], they were
matched in $g-r$ color to obtain the $ugriz$ absolute values, and
comparing them with the corresponding isochrones from DR10. This method
was designed by Schlesinger et al. (2012) for main-sequence stars, so we
used it to verify distances corresponding to stars with $\log g >$ 4.1,
ensuring they are dwarf stars. Another estimate of the distances for
main-sequence stars was obtained using the photometric parallax
relationship from Ivezic et al. (2008). We performed these tests for a
subsample of SDSS/SEGUE stars with $4.1< \log g < 4.4$, in our range of $T_{\rm eff}$ (16,760 objects). The correlations
of our estimates and these alternative approaches for dwarf stars are
shown in the top panels of Fig.~\ref{rdwarf_k}. They agree fairly
well, although our results tend to be slightly lower, with a mean
offset not greater than 0.6 kpc, and a dispersion $\sim$ 1 kpc. The
agreement is better with Schlesinger et al., probably because
the YREC isochrones were designed to work with
SEGUE data, while the photometric parallax relationship from Ivezic et
al. (2008) requires a conversion from the Johnson-Cousins system to
$ugriz$, increasing the uncertainties in the distance estimation (see Appendix B in Schlesinger et al. (2012)).

We also validated our methodology for giant stars by analyzing a sample
of 120 stars from Ramirez et al. (2013), who determined stellar
parameters for the stars in their sample for which Hipparcos parallaxes
were available. The comparison is shown in the bottom left panel of
Fig.~\ref{rdwarf_k} for stars with 2.5 $< \log g <$ 4.1 and effective temperature 5800 $< T_{\rm eff} <$ 6300 K. The agreement
is excellent. The formal comparison has
essentially a zero offset and a $\sigma$ $\sim$0.004 kpc.

Finally, we derived distances using the parameters of the SEGUE Stellar Parameter Pipeline (SSPP; Lee et al. 2008a,b, Allende Prieto et al. 2008, Smolinski et al. 2011, Lee et al. 2011) from the sample
of red giants analyzed by Xue et al. (2014), considering those with $2.5
< \log g < 3.5$ (493 objects), and comparing with their derived distance estimates
(although their effective temperatures are somewhat lower than the range
we are considering). The agreement is decent, but the dispersion ($\sim$
7 kpc) increases with distance, and our estimates tend to be somewhat
higher, as shown in the bottom right panel of Fig.~\ref{rdwarf_k}. 

The distances derived for our sample are listed in Tables \ref{tbl-2} and \ref{tbl-3}. It is worthwhile to note that although in some cases the estimated errors are significant, a more restrictive selection limited to relative errors smaller than $50\%$ does not change the results. For this reason we decided to include the entire sample. Our distance estimation code, {\tt get\_amrv.pro}, written in IDL, together with a file including a finely resampled version of the isochrones of Bertelli et al. is now publicly available on the web\footnote{http://hebe.as.utexas.edu/stools}.

\section{Results}

\subsection{[Fe/H] vs distance.}
\label{fehr_subsec}

\begin{figure*}
\center
\includegraphics[angle=90, scale=0.36]{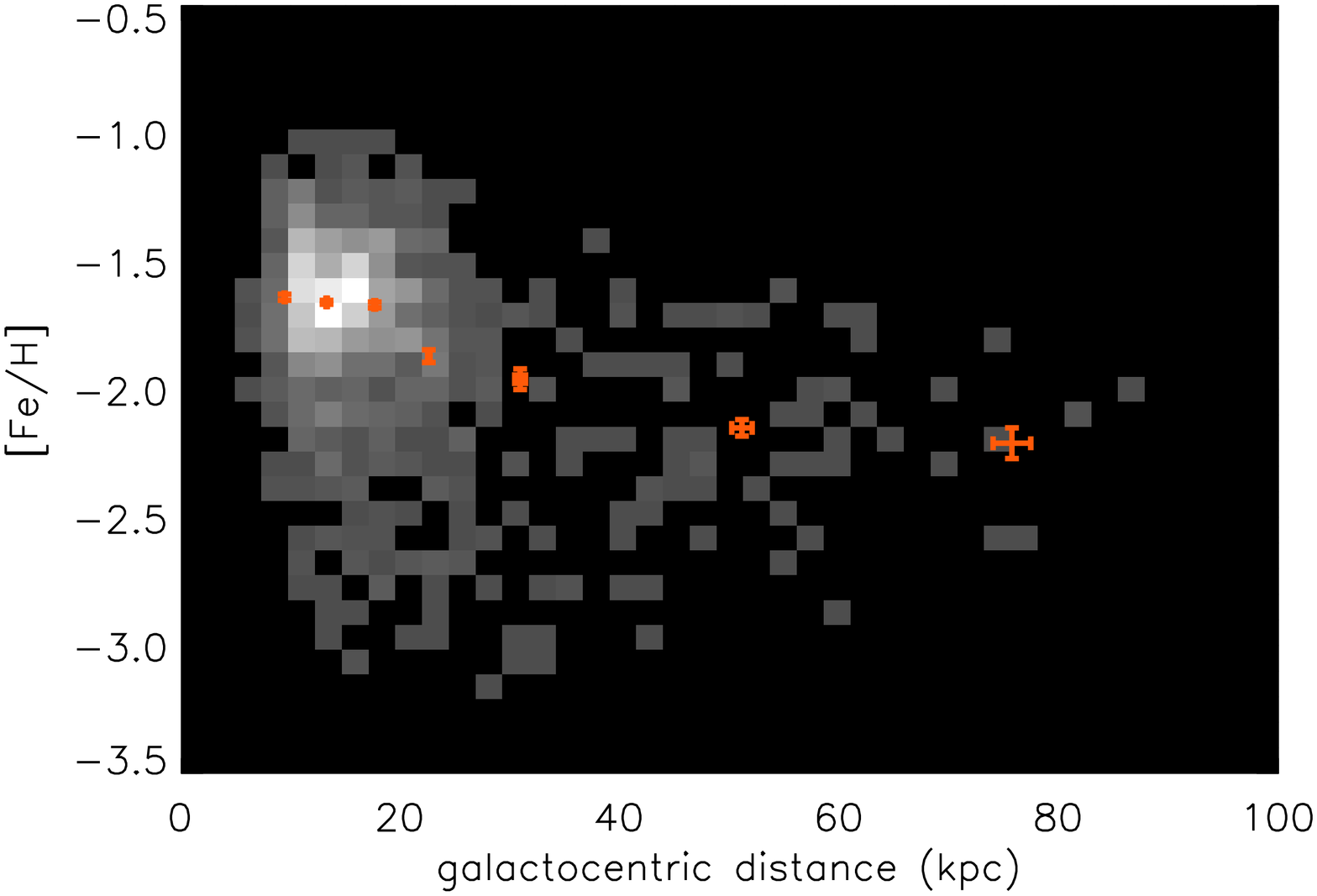} 
\includegraphics[angle=90, scale=0.36]{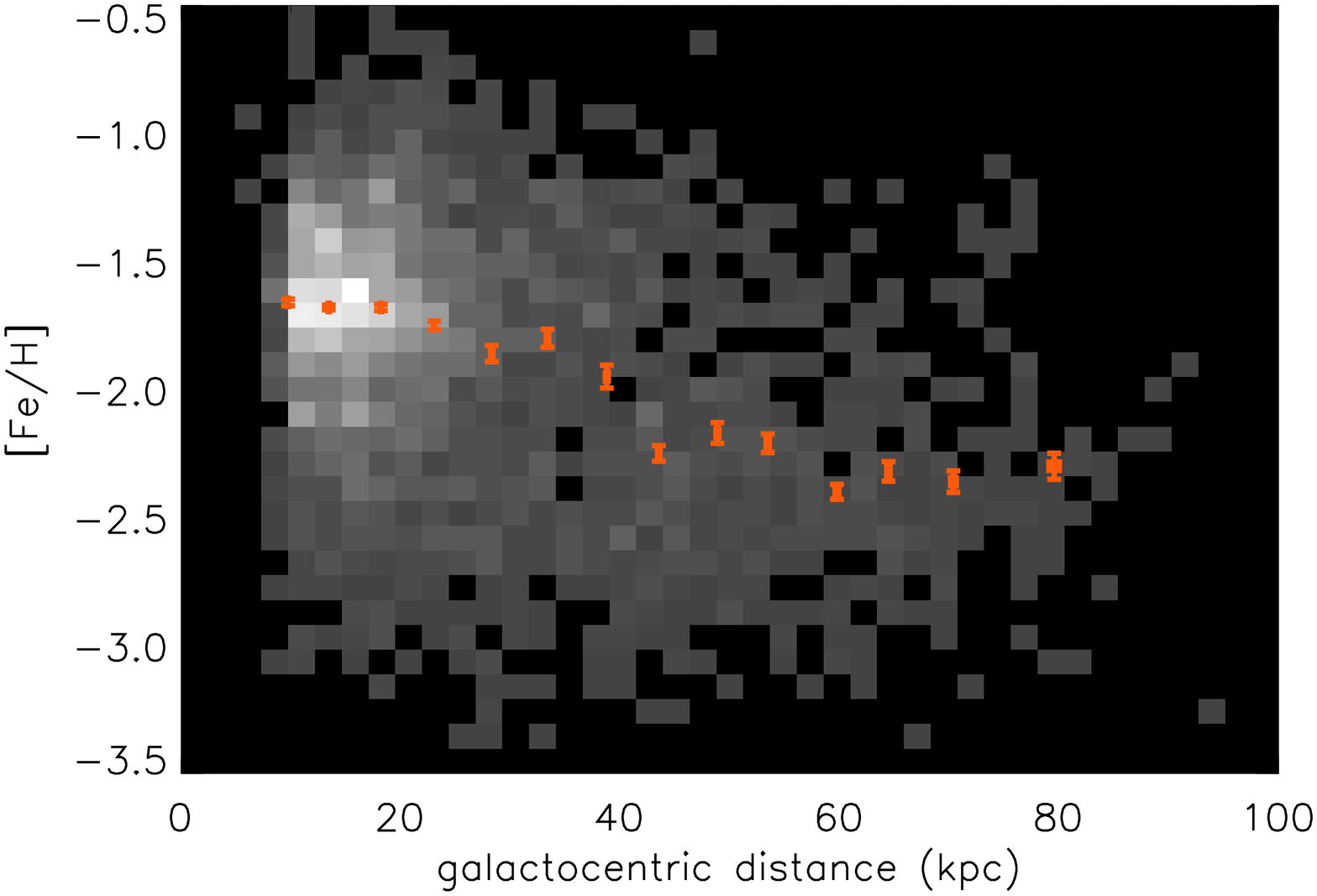}
\caption{Median [Fe/H] values, as a function of Galactocentric distance,
$r$, shown as a Hess diagram that displays the relative density of stars
for each bin of width 0.25 dex in [Fe/H] and 2.5 kpc in distance. The
orange points with their error bars summarize the median values of the analyzed set of data with
distance. The left panel corresponds to the BOSS F-type
spectrophotometric calibration stars, and the right panel to the full
SDSS/SEGUE sample (limited to the range described in Sect. 3.1, considering only stars with $|Z| > 5$ kpc). A clear
gradient in [Fe/H], decreasing with distance, is seen for both samples.}
\label{fe_R}
\end{figure*}

Using the abundances and distances for our sample of stars, we now
consider how [Fe/H] varies with distance from the center of the Galaxy,
$r$. We determined the median value of [Fe/H] in 5
kpc distance bins, except when this range does not include at least 50 stars, in which case
the bin was extended to cover the distance range needed to ensure this
minimum number of data points. We first calculated the median value of
the distance for stars in each bin. We selected only  estimates with errors lower than the dispersion known for the halo, 0.5 dex (Allende Prieto et al. 2006; Paper I). Assuming that the errors in our
estimates follow a Gaussian distribution, we calculated the error bars
as the median absolute deviation (MAD \footnote{MAD =
$median(|[Fe/H]_{i} - median_{j}([Fe/H]_{j})|)$}) divided by the square
root of the number of points. Figure~\ref{fe_R} shows the results of the
BOSS spectrophotometric calibration stars and the SDSS/SEGUE stars,
considering only those at a distance from the plane $|Z| > 5$ kpc to
avoid disk contamination. These samples comprise 1,109 objects from BOSS and 2,835 from SDSS/SEGUE. The stellar parameters, [Fe/H], [Ca/H] and [Mg/H] abundances and distance estimates are included in Tables \ref{tbl-2} and \ref{tbl-3}. It is clear, in both samples, that a gradient in the average value of
[Fe/H] exists in the halo -- at large distances the metallicity is clearly
lower than closer to the center. However, the trend is not identical in
the two cases. At the smallest distances, the iron abundances exhibit a
constant value of [Fe/H]$\sim -1.7$, begin to decrease at about 20 kpc,
and flatten out at about 40 kpc. The SDSS/SEGUE stars appear to
have slightly lower values of [Fe/H] at large distances.

\begin{figure*}
\center
\includegraphics[scale=0.8]{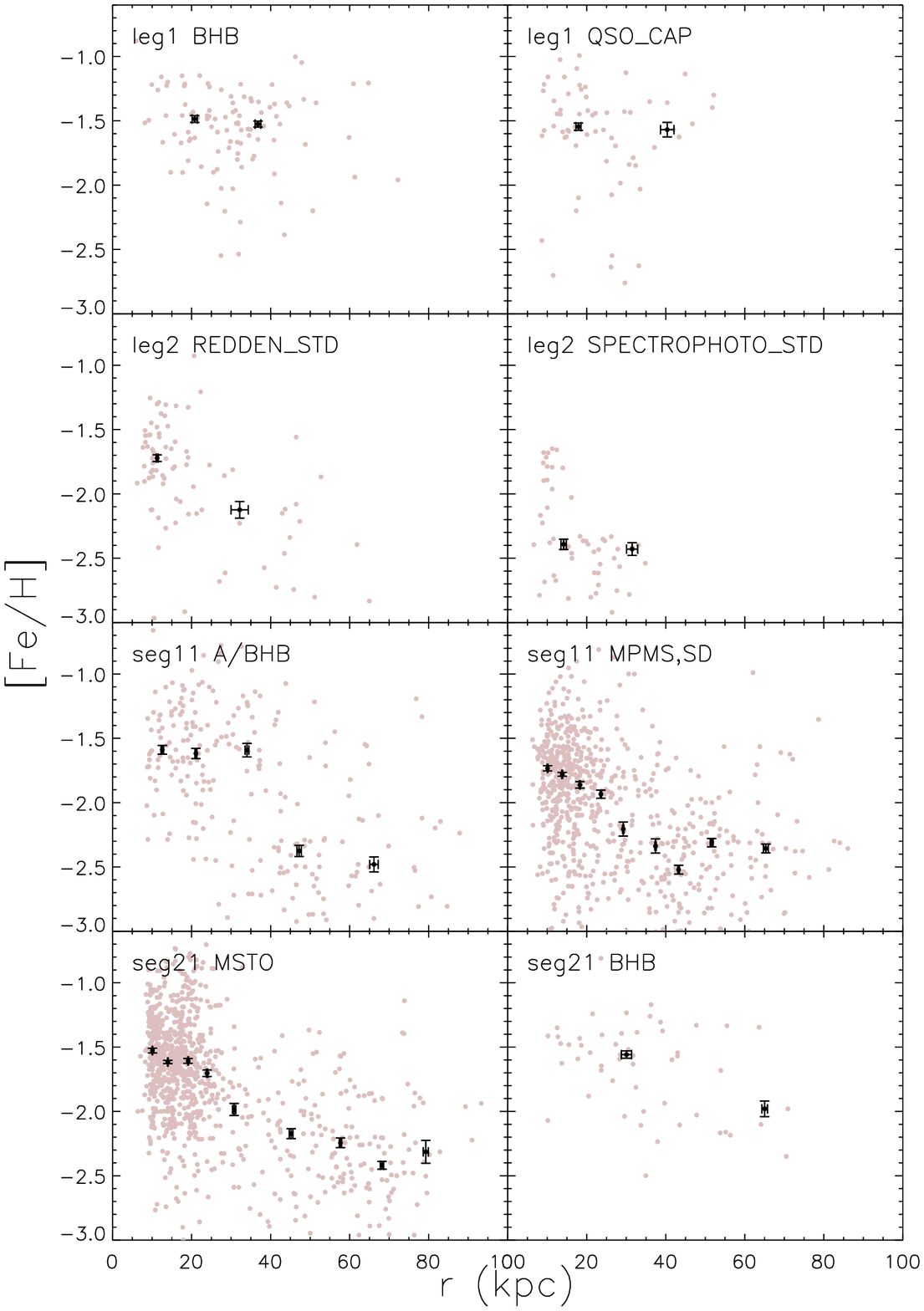}
\caption{Distribution of the median [Fe/H] vs. $r$, after splitting the
SDSS/SEGUE sample into the target-selection categories with more than 50
stars, located at distances from near the center to up to beyond 20 kpc
(www.sdss3.org/dr9/spectra/targets.php). Overall, the metallicity
decreases with distance the profile depends on the category
considered, however.}
\label{fe_distR}
\end{figure*}

As noted above, the SDSS/SEGUE sample comprises a number of different
target-selection categories, which may be responsible for the slightly
different behaviors seen in Fig.~\ref{fe_R}. To examine this in more
detail, we split our SDSS/SEGUE sample into its constituent target
categories \footnote{The legends of the panels in Fig.~\ref{fe_distR} refer to the program of the survey and whether the target selection belongs to science targets ('primary' target bits) or technical targets ('secondary' target bits):
'leg1'='Legacy-primary target'
'leg2'='Legacy-secondary target'
'SEG11'='SEGUE1-primary target'
'SEG12'='SEGUE1-secondary target'
'SEG21'='SEGUE2-primary target'
'SEG22'='SEGUE2-secondary target'.
For more information about the target selection and the different categories and their classification we refer to the webpage www.sdss3.org/dr9/spectra/targets.php}, shown in Fig.~\ref{fe_distR} (those including stars at
distances farther than 20 kpc, in order to be able to examine the trend
beyond this point). The most abundant categories are the MSTO,
metal-poor main-sequence (MPMS), and BHB stars. After applying the
same algorithm as before for each subsample, the results exhibit the same
decreasing trend of average [Fe/H] with increasing Galactocentric distance
(the shapes of the decreasing curves depend on the stars
considered). The median [Fe/H] value for BHB stars remains constant up to
$\sim$ 40 kpc, whereas for the MSTO and MPMS stars the
metallicity begins to decrease at $\sim$20 kpc, in agreement with the
BOSS sample. The SDSS/SEGUE stars reach slightly lower values of [Fe/H]
than the spectrophotometric standard stars of the BOSS sample.

\begin{figure*}
\center
\includegraphics[scale=0.58, angle=90]{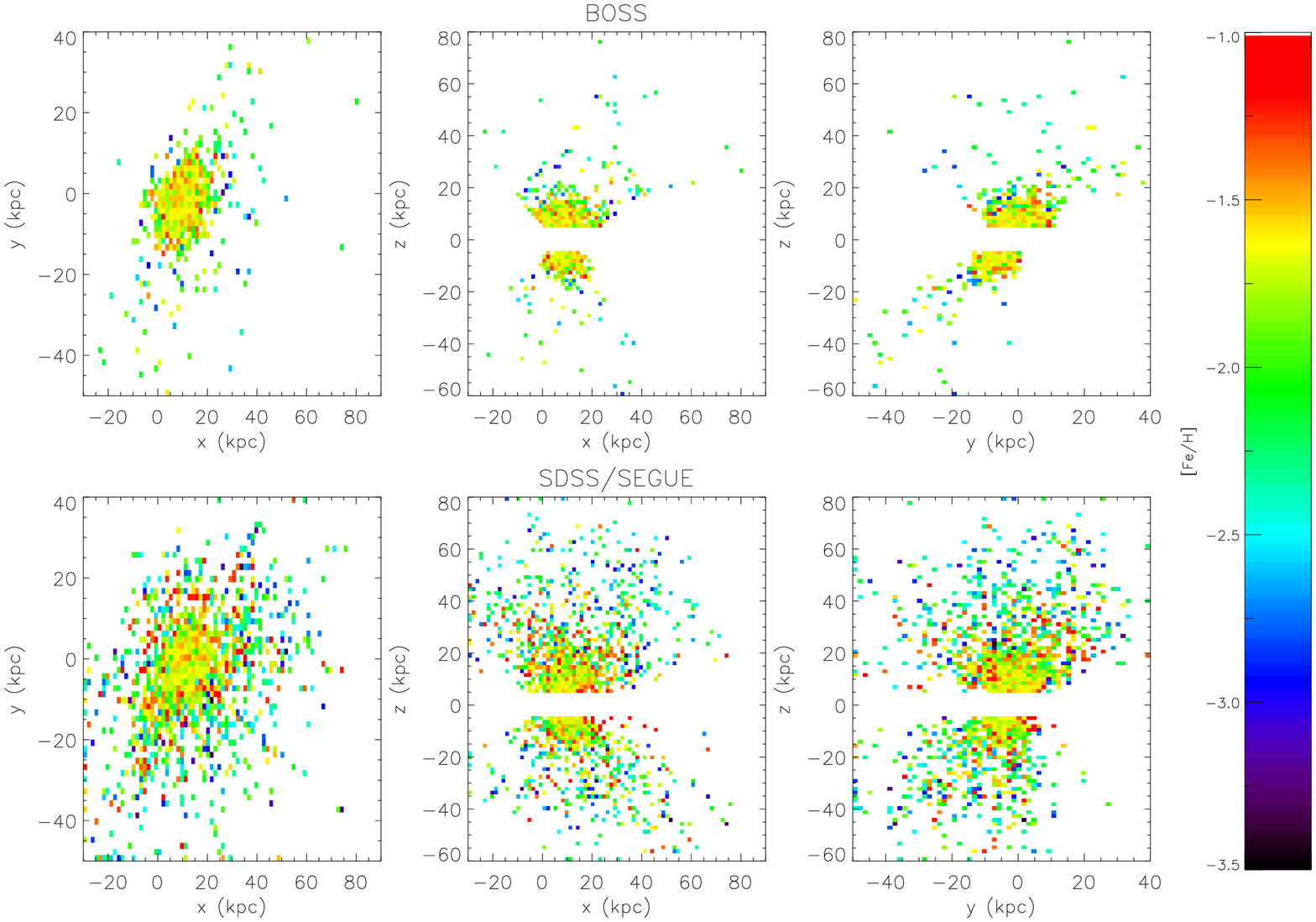}
\caption{Median values of [Fe/H], in bins of 1.5 kpc,
for the XY, XZ, and YZ planes of the Galaxy. There are regions
covered by the SDSS/SEGUE sample in the XZ
and YZ planes (the BOSS sample is shown in the upper row; the SDSS/SEGUE
sample is shown in the lower row) that are only poorly covered by the BOSS sample, in particular
at higher [Fe/H]. These stars primarily correspond to $r$ distances
between 20 and 40 kpc.}
\label{coord}
\end{figure*}

Figure~\ref{coord} displays the spatial distributions of the BOSS and
SDSS/SEGUE samples (in this reference system the Sun is at X$=+8$ kpc and Z
points towards the North Galactic Pole). Inspection of this figure
indicates that the SDSS/SEGUE stars cover a more extended range in the
different Galactic directions than the BOSS stars. Stars in the
transition region under discussion (20 kpc $< r < $40 kpc) were targeted
with far greater density by the SDSS/SEGUE sample.

\begin{figure}
\center
\includegraphics[scale=0.5]{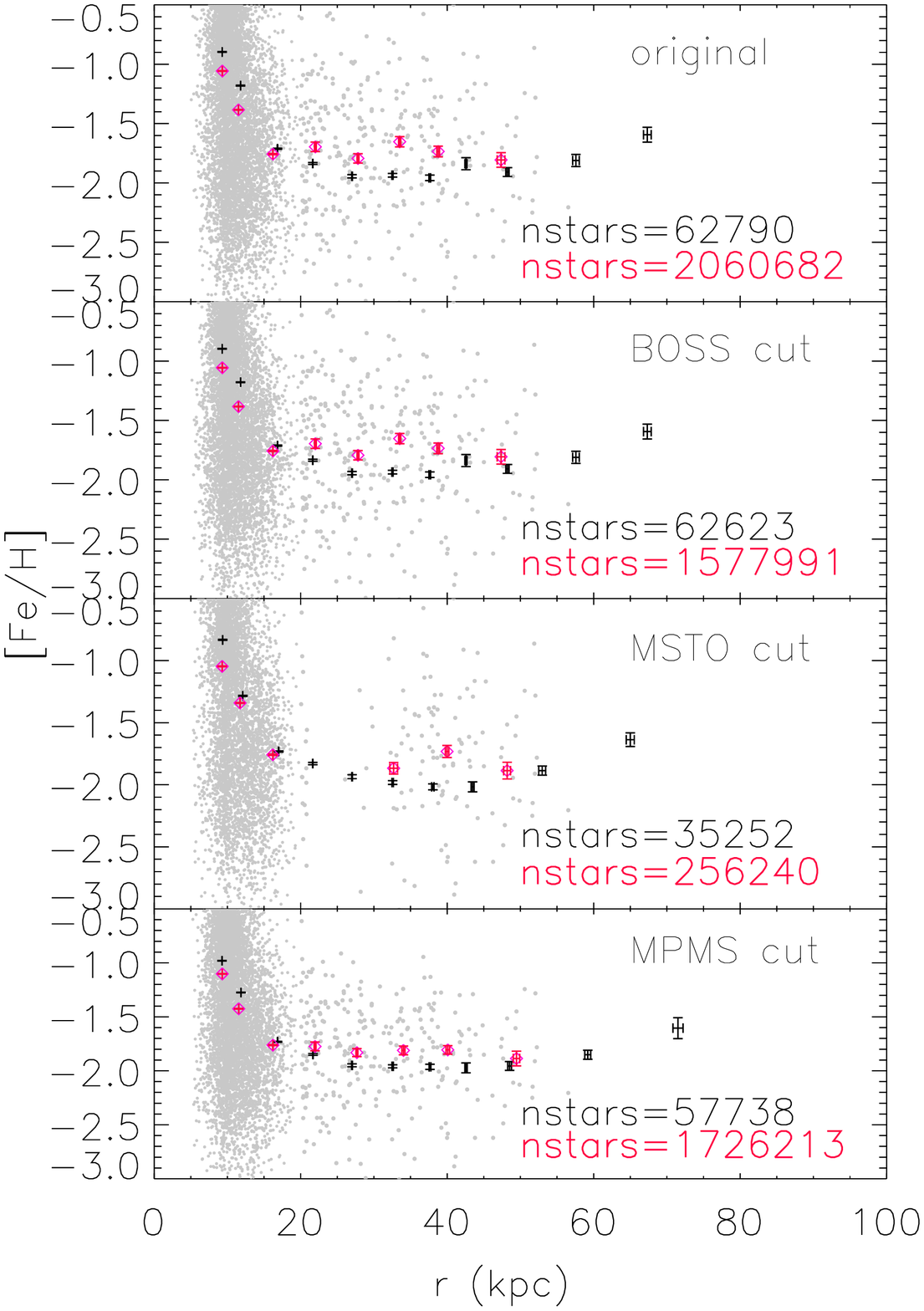}  
\caption{Besan\c{c}on simulation that reproduces stars in the directions
on the sky covered by BOSS. The top panel shows the original simulation; 
the other three panels correspond to the same stars for the BOSS spectrophotometric sample and the SEGUE MSTO
and MPMS target-selection cuts. In all cases we considered stars with
atmospheric parameters in the ranges of $T_{eff}$, $\log g$, and [Fe/H]
considered in this work. The plots show [Fe/H] with added noise vs distance from the Galactic center. Red diamonds represent the resulting median [Fe/H] values (with error bars) after applying our algorithm, considering the distance values from the model. Black dots correspond to the median [Fe/H] results, but considering
distances calculated from the stellar parameters and metallicity with
added noise. Although uncertainties in
distance estimates produce changes in the profile of [Fe/H] vs $r$,
it can be seen that neither the uncertainties in distance, nor
the stellar
parameters or the selection criteria considered in each case
generate a significant gradient.}
\label{color_cuts}
\end{figure}

To test whether our selection criteria in the stellar
parameters and colors could produce an artificial gradient, we have
considered a simulation of the behavior of [Fe/H] with $r$, based on
the Besan\c{c}on model of the Galaxy (Czejak et al. 2014), which assumes
no gradient in the halo. We consider a simulation of the Galaxy that
includes stars in several directions (those of BOSS plates), reaching
distances up to 50 kpc. For each star we have the stellar parameters and
metallicity values from the model, to which we added Gaussian noise (according to the
errors estimated from the dispersions obtained in Paper I when comparing
with the parameters from the SSPP: $\sigma_{T_{eff}}=70$~K,
$\sigma_{logg}=0.24$ dex, and $\sigma_{[Fe/H]}=0.11$ dex), and calculate
the distance following the same methodology as described above. Finally, we
selected stars with 5800~K $< T_{\rm eff} < 6300$~K, $2.5 < \log g <
4.4$ and $Z > 5$ kpc and applied the color restrictions that apply to the
BOSS spectrophotometric standard stars \footnote{$15.0 < r < 19.0$ and
$(((u-g) -0.82)^2 +((g-r)-0.3)^2+((r-i) -0.09)^2+((i-z) -0.02)^2)^1/2 <
0.08$}, MSTO SEGUE stars \footnote{$18.0 < g < 19.5$ and $0.1 < g-r <
0.48$}, MPMS SEGUE stars \footnote{$g < 20.3$, $0.4 < (u-g) < 0.7$, $0.2
< (g-r) < 0.7$ and $-0.7 < 0.91(u-g)+0.415(g-r)-1.28 < -0.25$} and the
BHB SEGUE stars \footnote{$15.5 < g < 20.3$, $-0.5 < g-r < 0.1$ and $0.8
< u-g < 1.5$}. Figure~\ref{color_cuts} shows the results after applying our algorithm, including the color cuts used for BOSS and SDSS/SEGUE, as well our distance binning. The high-metallicity values at r $< 20$ kpc indicate that, unlike the observations, stars from the thick-disk component of the model survive to the cut in $Z$ at 5 kpc. There were no stars that met the BHB color
restrictions, but Fig.~\ref{color_cuts} demonstrates that for the other three
cases these color cuts do not introduce an artificial gradient between 20-40 kpc. The uncertainties in our distance estimations cause oscillations in the median [Fe/H] values: a small decrease at r $> 20$ kpc and an increase at r $> 50$ kpc, but these variations are much smaller than what is observed in the analysis of the observed data. The last median value obtained at the largest distances deviates significantly with respect to the previous points and produces a positive gradient. However, this point is determined from few stars (26) and shows a wide dispersion in [Fe/H], unlike what is observed for real data.

\subsection{[Ca/H] and [Mg/H] vs distance.}

As described in Sect.~\ref{det_ab}, the calcium and magnesium
abundances are estimated by searching for the value of metallicity
among the theoretical spectra that best fit transitions of each
element. From this value the [X/H] abundance is obtained,
taking into account the relation between the iron and $\alpha$-element
abundances adopted in generating the synthetic spectra (see Paper I).

\begin{figure*}
\center
\includegraphics[width=6.5cm]{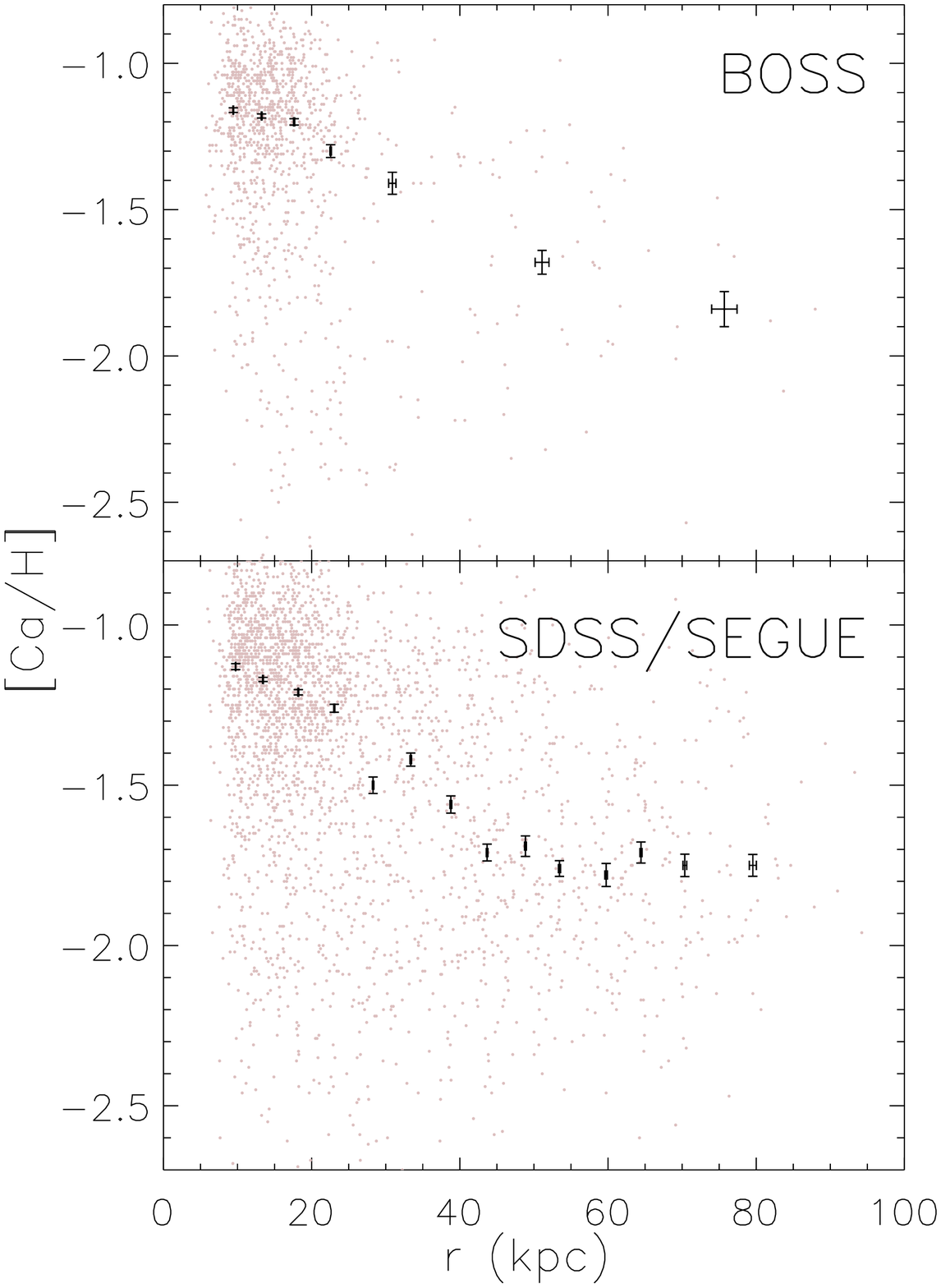}
\includegraphics[width=6.5cm]{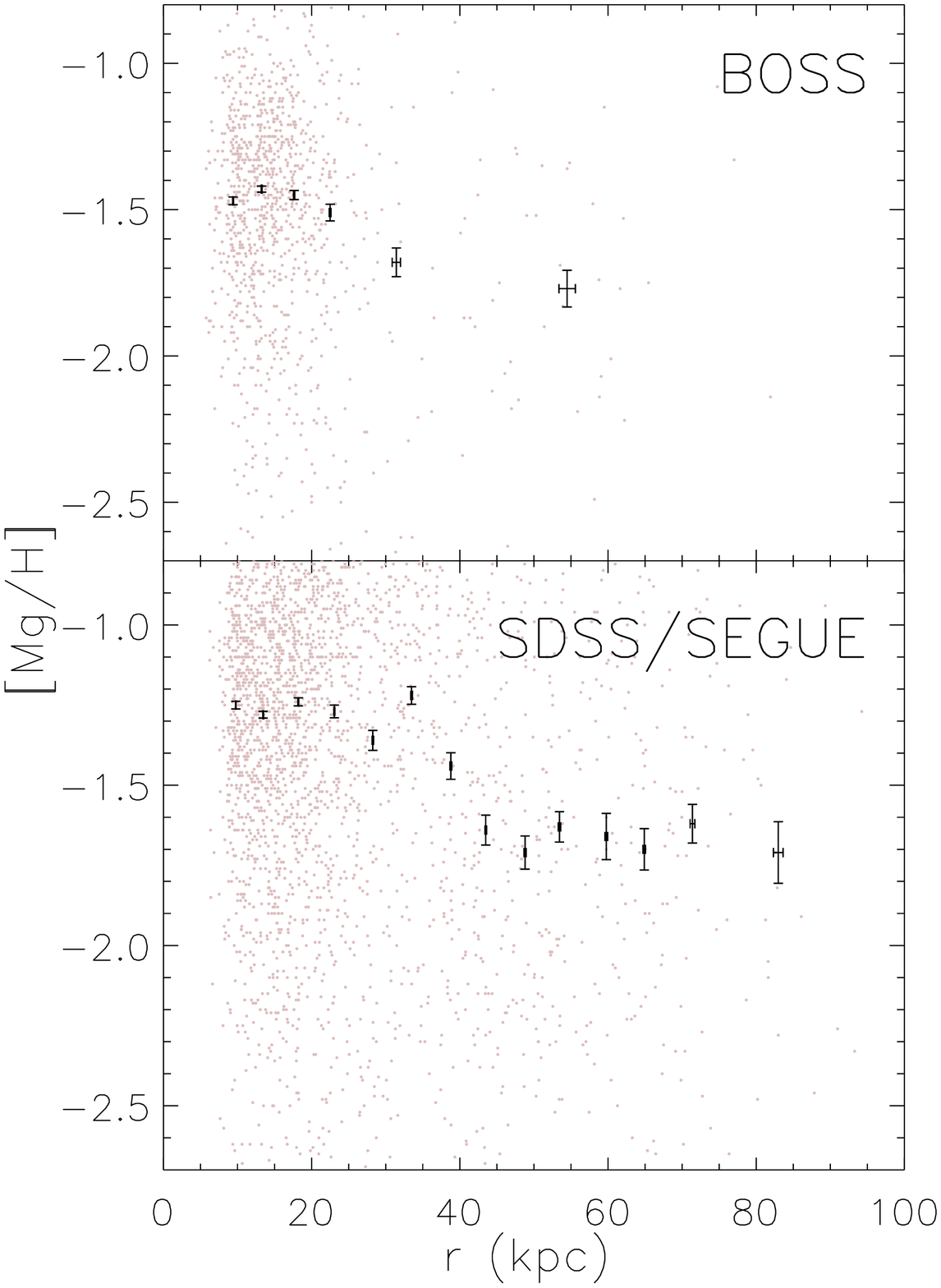}
\caption{Median [Ca/H] and [Mg/H] abundances as a function
of Galactocentric distance, $r$. The upper panels correspond to the BOSS
spectrophotometric sample, while the lower panels are for the SDSS/SEGUE
sample.}
\label{ca_distR}
\end{figure*}

Figure~\ref{ca_distR} shows the observed [Ca/H] and [Mg/H] trends with
Galactrocentric distance. These trends reinforce the conclusion that
there exists a chemical gradient in the halo. Although there are some differences between the samples, stars at
large distances exhibit lower [Ca/H] and [Mg/H] values. The trend observed
for calcium is to decrease from [Ca/H]$\sim-1.2$ to [Ca/H]$\sim-1.8$ at the
largest distances. The larger number of SDSS/SEGUE stars permits a greater
number of [Ca/H] bins for the more distant stars, and
from these data we conclude that there is a flat trend beyond 40 kpc.

For magnesium, the analysis of the BOSS sample shows a
decreasing trend of [Mg/H] values that are $\sim$ 0.2 dex lower than for SDSS/SEGUE stars at distances shorter than 40 kpc.
It exhibits a constant value up to $r\sim20$ kpc, and the same flat trend
at the largest distances see for [Ca/H], at around the same median value.

\subsection{[Ca/Fe] and [Mg/Fe] ratios.}

\begin{figure*}
\center
\includegraphics[width=6.5cm]{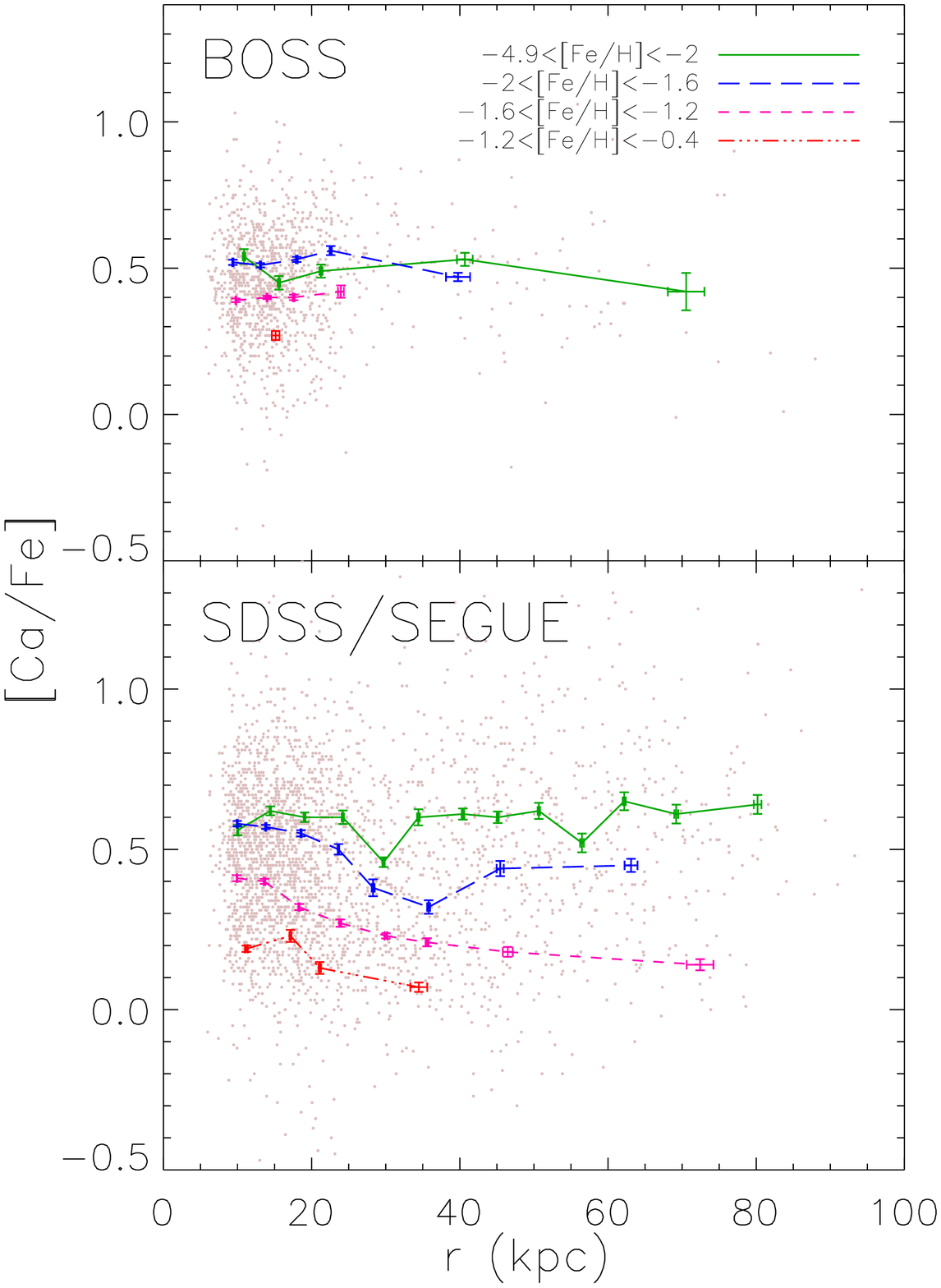}
\includegraphics[width=6.5cm]{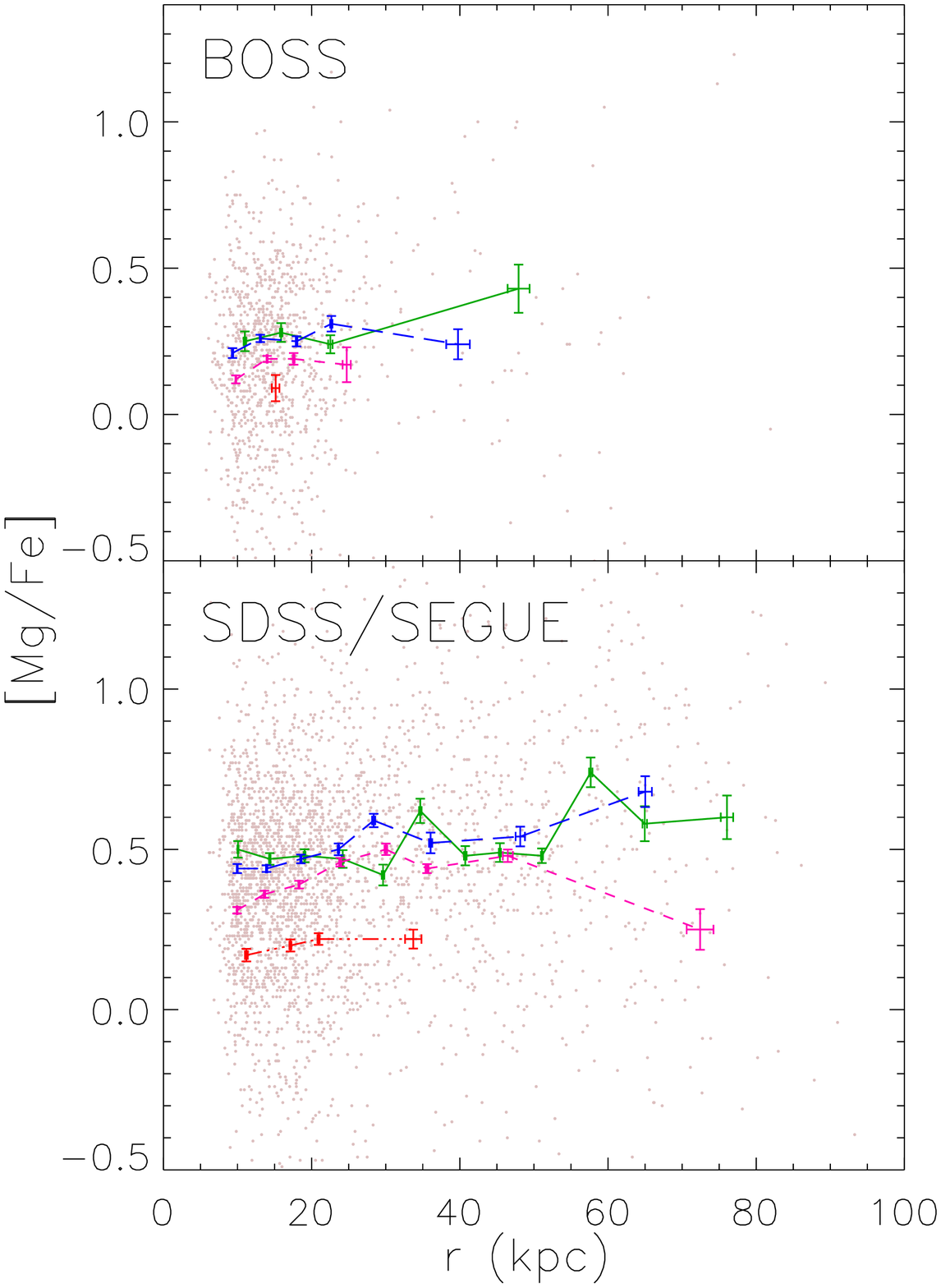}
\caption{Median [Ca/Fe] and [Mg/Fe] abundance ratios as a function of
the Galactocentric distance, $r$.  The colored lines indicate different
ranges of [Fe/H]. The trends of these ratios with distance depend on the range of metallicity and the element under consideration. }
\label{cafe_r}
\end{figure*}

Finally, we examine the [Ca/Fe] and [Mg/Fe] abundance ratios resulting
from the [Ca/H], [Mg/H] and [Fe/H] estimates. Figure~\ref{cafe_r}
presents the trends of these ratios with Galactocentric distance over
four different ranges of metallicity. The observed trends are not
identical for the BOSS and SDSS/SEGUE samples, but the results are fairly
consistent. For this analysis, we note the small number statistics as a result of dividing the sample into several metallicity bins, especially for the BOSS sample. The figure clearly shows that metal-rich stars are missing at large distances. 

For [Ca/Fe], both
samples reveal that this ratio tends to become higher as the abundance
of [Fe/H] decreases. Although the analysis with BOSS stars returns
constant trends, SDSS/SEGUE stars exhibit a decreasing behavior with
Galactocentric distance, except at the lowest metallicities, where the
trend appears flatter. Similarly, the [Mg/Fe] estimates are lower for
stars over the range $-1.2 <$ [Fe/H] $< -0.4$ than at lower values of
[Fe/H]. At distances larger than 20 kpc it is not possible to distinguish a trend
with metallicity. There is a subtle increasing trend with distance,
mainly at $r < 40$ kpc and [Fe/H] $> -2$. For this abundance ratio, it
is difficult to analyze the outer region of the halo because
of the large
error bars and the high dispersion in [Mg/H] estimates for more
metal-poor stars. 

\begin{figure*}
\center
\includegraphics[width=6.5cm]{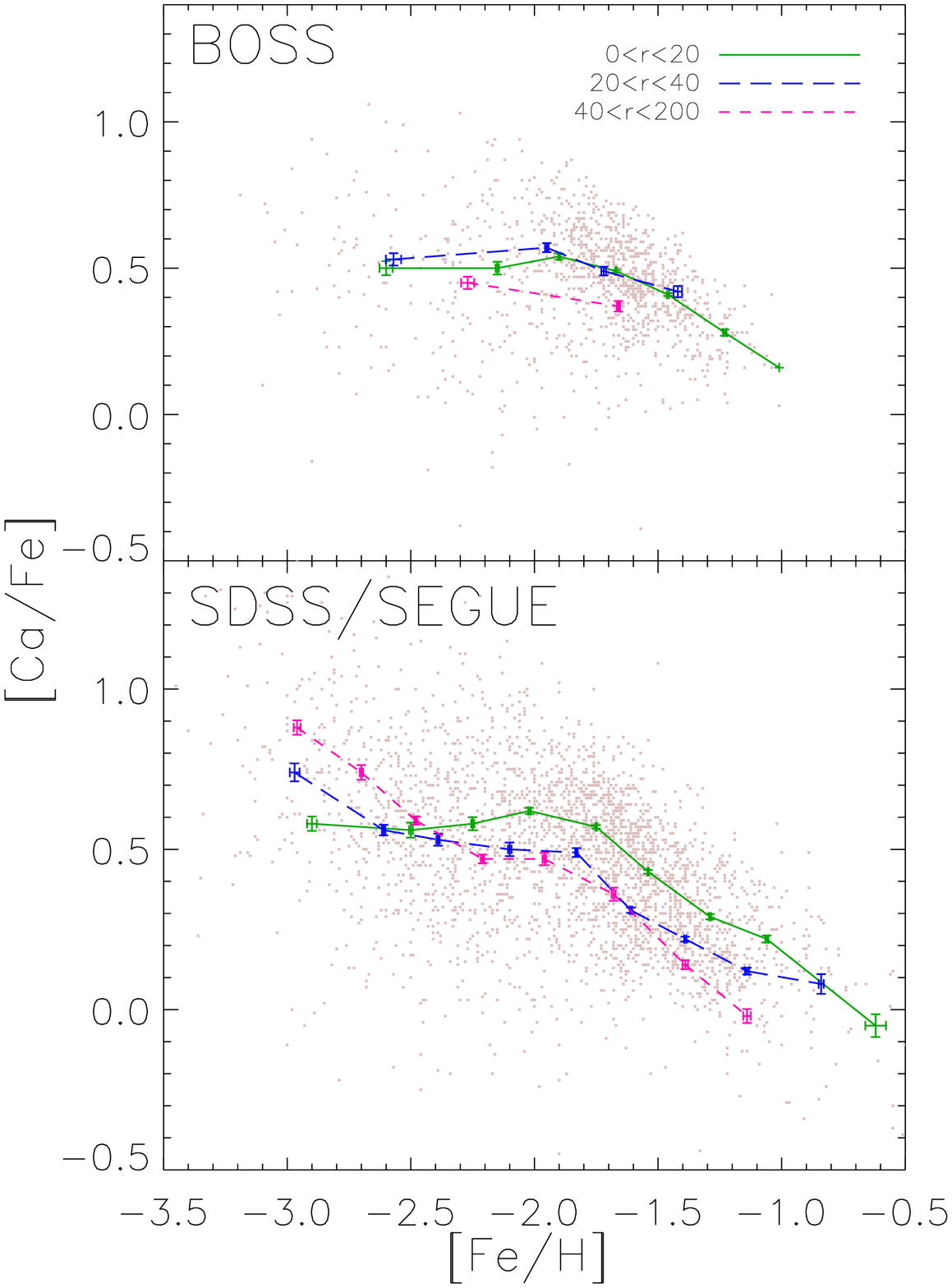}
\includegraphics[width=6.5cm]{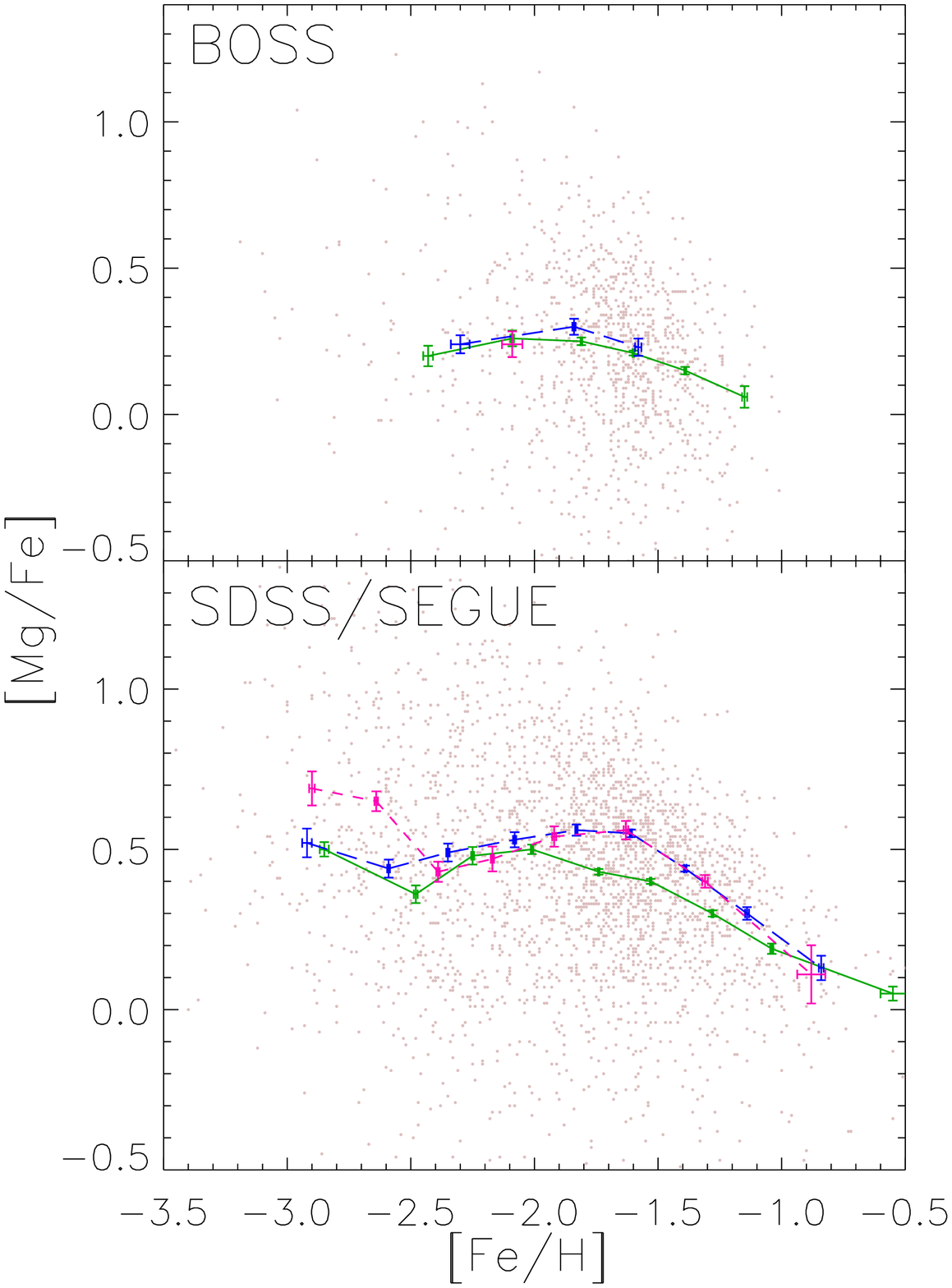}
\caption{Median [Ca/Fe] and [Mg/Fe] as a function of metallicity, [Fe/H].
The colored lines indicate different ranges of Galactocentric distance, $r$. The [$\alpha$/Fe] ratio 
increases as the metallicity decreases to [Fe/H] $\sim -2$, after with the trend
flattens. Note the steeper trend observed for the SDSS/SEGUE stars at [Fe/H]
$<-2.5$ at the largest distances.}
\label{cafe_met}
\end{figure*}

Figure~\ref{cafe_met} compares the [Ca/Fe] and [Mg/Fe] as a
function of [Fe/H], split into three ranges of distance: $0 < r < 20$
kpc, $20 < r < 40$ kpc, and $r > 40$ kpc. The median [Fe/H] values were
calculated in bins of 0.25 dex, or wider if the number of stars in this
range does not reach at least 50. The BOSS sample exhibits a similar
behavior for both abundance ratios: lower [$\alpha$/Fe] ratios
for the most distant stars with [Fe/H]$<-2$, without significant differences between
stars at $r < 40$ kpc. At the lowest [Fe/H], [Ca/Fe] increases significantly with
distance. For the SDSS/SEGUE sample (which comprises more stars), the
most metal-poor stars exhibit higher [$\alpha$/Fe] than for the
metal-rich stars, increasing with distance. The resulting [Ca/Fe] 
reveal a decreasing trend with metallicity from [Fe/H] $\sim-2.0$ to the highest values considered. At [Fe/H] lower than $\sim -2$, [Ca/Fe] remains almost
constant for nearer stars, but increases steeply as the metallicity
decreases for distant stars. The trend of [Mg/Fe] with [Fe/H] is very
similar for stars at $r< 20$ kpc, but more distant stars exhibit 
a steeper trend with metallicity and higher values than nearer stars (although the metal-poor tail suggests the
same trend with distance that is observed for Ca).

\subsection{Effects due to systematics.}

We now consider whether the presence of systematic offsets in our stellar parameter estimates might be introducing artificial effects in the determination of the abundances.  In Paper I, we found that both [M/H] and surface gravity were systematically underestimated in low-metallicity giants. This effect was clearly seen in stars with $\log g <$ 2.5 in the globular cluster M 13. There is no obvious evidence of a systematic  trend with $\log g$ for our [Mg/H] and [Fe/H] in M3. Conversely, [Ca/H] shows the same trend detected for [M/H] as in Paper I, as expected given that at very low metallicity the Ca II lines dominate the [M/H] determination from the entire spectrum. We trimmed our sample by keeping only stars with $\log g >$ 2.5 to prevent systematic errors.

The decision to cut at $\log g$=2.5 was made qualitatively.  To ensure the adequacy of this choice, we verified in a sample of SDSS/SEGUE stars at $2.5 < \log g < 3$ that an increment of 0.3 (the offset detected in Paper I for stars at $\log g < 3$ comparing with $\log g$ estimates from the SSPP) implies variations lower than 0.1 in the abundances of the three elements. The [Fe/H], [Ca/H] and [Mg/H] changes as a function of distance show higher variations. Consequently, they cannot be due to systematics. As described in Paper I, the comparison of our SDSS/SEGUE $T_{\rm eff}$ estimates with those from the SSPP agrees very well and has no systematic offsets.We compared more than 90 stars that match our SDSS/SEGUE and BOSS samples. We found an offset in temperature of -80$\pm$5 K that increases at lower temperatures. This is much too small an offset to compromise our conclusions.

\section{Conclusions.}

We have performed an analysis of SDSS optical stellar spectra and
derived elemental abundances for magnesium, calcium, and iron in individual
stars in the halo of the Milky
Way. We split these halo stars into two samples depending on whether they have been observed before (SDSS observations up to fall 2009) or after the update of the SDSS spectrographs (BOSS; Eisenstein et al. 2011).

For both samples, we observe a clear decreasing trend
of metallicity with distance from the Galactic center. The decrease becomes obvious at $r \sim 20$ kpc, in fair agreement with earlier studies (Carollo et al. 2007, 2010; de Jong et al. 2010; Beers et al.
2012; Chen et al. 2014). The investigations of Carollo et al. and Beers et al. obtained their results from a spectroscopic analysis of nearby halo stars identified from their
kinematics, while the work of de Jong et al. was based on photometry. Our analysis (and also that of Chen et al.), however, is based on
halo stars observed spectroscopically in situ over a broad range of distances.

The individual abundances of the two $\alpha$-elements Ca and Mg confirm the chemical gradient in the halo, which flattens out beyond
$r \sim$ 40 kpc. The trends inferred from this work for the [Ca/Fe] and [Mg/Fe] abundance
ratios as a function of Galactocentric distance exhibit differences that depend on the metallicity range and the element considered: the ratios are roughly flat at metallicities [Fe/H]$< -2$ for Ca, but decrease with distance at higher metallicities. [Mg/Fe] shows the opposite trend: it increases with distance regardless of metallicity. However, the interpretation of [Mg/Fe] should be considered as preliminary, given the high dispersion detected in our [Mg/H] estimates and the apparent oscillations as a function of Galactocentric distance. Stars with [Fe/H] $< -2$ exhibit the highest values of the
$\alpha$/Fe abundance ratio in the halo. The
[Ca/Fe] and [Mg/Fe] decrease with the iron abundance, although there is a
range in metallicity at which these ratios remain constant.

For the most metal-poor stars, a range where $\alpha$/Fe ratios have not been studied in detail before, our analysis shows higher ratios than in more metal-rich stars. These trends are steeper at larger Galactocentric
distances, suggesting different $\alpha$-enrichment histories for the inner and outer parts of the halo.

The hierarchical assembly predicted by the $\Lambda$CDM model for the formation of massive galaxies implies the total or partial disruption of the subgalactic systems from which the Milky Way halo would have been formed. The mixing of disrupted protogalaxies would form a smooth halo, and the surviving cores would appear as overdensities, with particular kinematical and chemical properties.

Nissen \& Schuster (2010) reported evidence for two different [Mg/Fe] and
[Ca/Fe] levels for halo stars in the solar neighborhood,  with
kinematics that indicate a low-$\alpha$ outer halo and a high-$\alpha$
inner halo. Their analysis is limited to stars at $-1.6 <$ [Fe/H] $< -0.4$. In this metallicity range, our Ca results agree with the assumption of distant stars with lower
[$\alpha$/Fe]; however, at the lowest metallicities we find no evidence of such a trend. On the contrary, we measure higher [Mg/Fe] abundance ratios for the most distant stars at all metallicities. 

Tissera et al. (2013, 2014) performed an analysis of six simulated
halos of Milky Way-like galaxies, taking
into account supernova feedback, metal-dependent cooling, a prescription
for star formation, and a multiphase model for the gas component. They
found that  outer-halo populations (defined using an energy
criterion) are characterized by having low
metallicity and high $\alpha$-element enrichment. Their inner-halo populations comprise debris stars and
disk-heated stars, the latter with lower $\alpha$-element
enhancements and more gravitationally bounded than the former. They
have also considered that some (in particular more massive) subgalactic
fragments (or satellites) can survive longer into the potential well of
their parent galaxies, while retaining gas from which new stars can be
formed (referred to as endo-debris stars), with low values of [Fe/H], but
generally a lower $\alpha$-element enrichment than debris stars. The
disk-heated stars are mostly formed in the disk and later scattered to the halo, with low $\alpha$-element
enhancement due to contributions from Type Ia supernova yields. The simulations
showed about 15\% of accreted stars with higher
$\alpha$-element enhancement and larger dispersion, consistent with the
properties of a thick-disk component. For five of the six cases, they
determined their halo transition radii (the crossover point between regions
that reflect dominant contributions from the inner-halo populations and
outer-halo populations, respectively) were located at about 15-20 kpc from the Galactic
center, similar to our results. One of their simulated halos exhibited a
halo transition region located at $\sim 40$ kpc, similar to that
reported for M31 by Gilbert et al. (2014).

Our new sample is not the only one available to explore the properties
of metal-poor stars in the distant halo. Chen et al. (2014) analyzed
spectra of in situ halo red giants from the SDSS ninth data release
(DR9; Ahn et al. 2012). Their study revealed a metallicity distribution function
that transitioned from unimodal to bimodal at distances 20 $< |Z| <$ 25
kpc and 35 $< r < $45 kpc, with peaks at [Fe/H] $\sim-1.6$ and [Fe/H]
$\sim-2.3$, the first associated with an inner-halo population and the
second with an outer-halo population. These values are similar to the
median metallicity that we observe in the inner and outer regions of the
Galactic halo. They also found the inner-halo population to dominate at
4 $< |Z| <$ 10 kpc and located the transition point between the
inner-halo and outer-halo populations at r $\sim 35$ kpc.

The predictions of Tissera et al. (2013,2014) agree with our results that the most distant stars have higher $\alpha$/Fe ratios than the inner region. In contrast, at higher metallicities our [Ca/Fe] are lower for more distant stars than nearer ones, in agreement with Nissen \& Schuster, and in contradiction with the simulations of Tissera et al.
The variation of [$\alpha$/Fe] as a function of [Fe/H] that we find in the outer halo differs from the common image of a plateau caused by the $\alpha$-enrichment due to SNII without the contribution from SNIa. The [$\alpha$/Fe] curve as a function of the iron abundance in the outer regions shows a decline from the lowest [Fe/H], remains constant over a certain range, and decreases again.

This curve is the average of stars that are placed at the farthest distances in the halo. As we mentioned above, the accretion of subgalactic systems plays a role in the formation of the halo in the $\Lambda$-CDM model, and they are expected to be smaller structures than the resulting galaxy. There is observational evidence that low-mass systems show a decrease of [$\alpha$/Fe] at lower [Fe/H]. As a result of the lack of mass, they are not expected to sustain the formation of massive stars for a long time. In addition, winds caused by the first SN are likely to remove the gas of the system, which reduces the star formation rate. This implies that SNIa start to enrich the interstellar medium at a lower Fe/H than in more massive systems (Venn et al. 2004).

If the subsystems accreted by our Galaxy would have covered a range of masses, their $\alpha$/Fe vs Fe/H curves would show a decrease at different Fe/H values. The average curve would be drawn by the superposition of all of them. From the results reported in this work, the lowest metallicities appear to be dominated by stars that belonged to very low-mass subsystems with the contribution of SNIa at very low [Fe/H]. On the other hand, an IMF biased to high masses would lead to higher $\alpha$/Fe ratios (McWilliam 1997). The different slopes observed for [$\alpha$/Fe] at the lowest metallicities, with higher levels at the outer most region, may be a signature of a different IMF, skewed to higher mass progenitors, in the most distant regions of the halo.

The inner parts of the halo are expected to be dominated by stars formed within the virial radius of the Galaxy, although a fraction of debris stars could also contribute, as suggested from the
simulations of Tissera et al. For the nearest region the [$\alpha$/Fe] trend with [Fe/H] that we observed is consistent with a higher star formation rate than outer regions.  

Tissera et al. (2014) concluded that their simulated halos that
exhibited a mild metallicity gradient had significant contributions from
relatively more massive subgalactic fragments. On the other hand, the
outer-halo populations that originated from low- or intermediate-mass
systems would produce flatter metallicity profiles with distance. The
abundance gradient observed in our study of the Milky Way halo,
which flattens at
large distances, appears to be the result of more massive subgalactic
fragments contributing to the inner region, whereas the outer region
would be formed from the accretion of systems with lower mass
and lower metallicity, leading to a flat metallicity
profile with distance.

SDSS optical observations will continue at least
for another six years, and as result,
a larger sample of F-type turn-off halo stars will soon be available, enabling further investigation and expansion of the results described in this study.

\begin{acknowledgements}

T.C.B. acknowledges partial support for this work from grants
PHY 08-22648; Physics Frontier Center/{}Joint Institute or Nuclear
Astrophysics (JINA), and PHY 14-30152; Physics Frontier Center/{}JINA
Center for the Evolution of the Elements (JINA-CEE), awarded by the US
National Science Foundation. AR acknowledge the support of the French Agence Nationale de la
Recherche under contract ANR-12-BS05-0015. BGM simulations
were executed on computers from the Utinam Institute of the
Universit\'e de Franche-Comt\'e, supported by the R\'egion de
Franche-Comt\'e and Institut des Sciences de l'Univers (INSU). Y.S.L. acknowledges partial support by the 2014 research fund of Chungnam National University.

Funding for SDSS-III has been provided by the Alfred P. Sloan Foundation,
the Participating Institutions, the National Science Foundation, and the U.S.
Department of Energy Office of Science. The SDSS-III web site is http://www.sdss3.org/.

SDSS-III is managed by the Astrophysical Research Consortium for the
Participating Institutions of the SDSS-III Collaboration including the
University of Arizona, the Brazilian Participation Group, Brookhaven National Laboratory,
University of Cambridge, Carnegie Mellon University, University of Florida,
the French Participation Group, the German Participation Group, Harvard University,
the Instituto de Astrofisica de Canarias, the Michigan State/Notre Dame/JINA
Participation Group, Johns Hopkins University, Lawrence Berkeley National Laboratory,
Max Planck Institute for Astrophysics, Max Planck Institute for Extraterrestrial Physics,
New Mexico State University, New York University, Ohio State University,
Pennsylvania State University, University of Portsmouth, Princeton University,
the Spanish Participation Group, University of Tokyo, University of Utah,
Vanderbilt University, University of Virginia, University of Washington,
and Yale University.

\end{acknowledgements}

\addtocounter{table}{1}

\longtab{1}{

\begin{longtable}{ccc}

\caption{List of lines used to determine chemical abundances.\label{tbl-1}}\\
\hline\hline
ionization state & $\lambda$ & EW(m\AA)\footnotemark\\
\hline
\endfirsthead
\caption{continued.}\\
\hline\hline
ionization state & $\lambda$ & EW(m\AA) \\
\hline
\endhead
\hline
\endfoot
\footnotetext{Approximated value of the equivalent width for each line calculated in the generation of a synthetic spectra at $T_{\rm eff}$ = 6000 K, $\log g$=4.0 dex and [Fe/H] = -1 dex using the SYNSPEC code (Hubeny et al. (1985); http://nova.astro.umd.edu/Synspec49/synspec.html).}
Fe     &      4885-5000    &   $100\%$ used\footnotemark\\
\hline
\footnotetext{This quantity indicates the percentage of the selected spectral region used in calculating the $\chi^{2}$ parameter, in order to mostly include lines that belong to the element whose abundance is desired to be measured, and avoid features due to other elements.}
Fe I &     4885.43   & 20 \\
Fe I  &    4886.33 &   19 \\
Fe I  &    4887.19 &   12 \\
Fe I  &    4889.00 &   35 \\
Fe I  &    4889.10 &   11 \\
Fe I  &    4890.75 &   83 \\
Fe I  &    4891.49 &   90 \\
Fe I  &    4903.31 &   66 \\
Fe I  &    4910.32 &   37 \\
Fe I  &    4910.56 &   37 \\
Fe I  &    4918.95 &   52 \\
Fe I  &    4918.99 &   85 \\
Fe I  &    4920.50 &    95 \\
Fe II &    4923.93 &  102 \\
Fe I  &    4924.77 &   43 \\
Ba II &    4934.08 &   98 \\  
Fe I  &    4938.81 &   66 \\
Fe I  &    4939.69 &   58 \\
Fe I  &    4946.39 &   41 \\
Fe I  &    4957.30 &   87 \\
Fe I  &    4957.60 &   98 \\
Fe I  &    4957.68 &   48 \\
Fe I  &    4966.09 &    57 \\
Fe I  &    4978.60 &    33 \\
Fe I  &    4978.69 &   45 \\
Fe I  &    4982.50 &   67 \\
Fe I  &    4983.25 &   56 \\
Fe I  &    4983.85 &    60 \\
Fe I  &    4985.25 &   46 \\
Fe I  &    4985.55 &   58 \\
Fe I  &    4994.13 &    66 \\ 
\hline
 Fe  &      5220-5350   &     $100\%$ used \\
\hline 
Fe I   &   5226.86  &  81 \\   
Fe I  &    5227.15 &   80\\
Fe I  &    5227.19 &  103\\
Fe I  &    5229.84 &   55\\
Fe I  &    5229.87 &   53\\
Fe I  &    5232.94 &   93\\
Fe II &    5234.62 &   74\\
Fe I  &    5242.49 &   49\\
Fe I  &    5250.65 &   59\\
Fe I  &    5253.46 &   20\\
Fe II &    5254.93 &   19\\
Fe I  &    5254.95 &   22\\
Si I  &    5258.84 &   47 \\
Si I  &    5260.66 &   12\\
Ca I  &    5261.70 &   20\\
Ca I  &    5262.24 &   35\\
Fe I  &    5263.31 &   60\\
Ca I  &    5265.56 &   57\\
 Fe I  &    5266.56 &   85\\
 Fe I  &    5269.54 &  119\\
Ca I  &    5270.27 &   69\\
Fe I  &    5270.36 &   96\\
Fe I  &    5273.16 &   57 \\
Fe I  &    5273.37 &   35 \\
Fe II &    5276.00 &   80\\
Fe I  &    5281.79 &   67\\
Fe I  &    5283.62 &   74\\
Fe II &    5284.11 &   26 \\
Cr I  &    5296.69 &   45\\
Cr I  &    5298.27 &   59\\
Fe I  &    5302.30 &   64 \\
Fe I  &    5307.36 &   37\\
Fe II &    5316.61 &   85\\
Fe II &    5316.78 &   24\\
Fe I  &    5324.18 &   87\\
Fe II &    5325.55 &   50\\
Fe I  &    5328.04 &  116\\
Cr I  &    5328.53 &   44\\
Fe I  &    5328.53 &   89\\
Fe I  &    5332.90 &   45\\
Si I  &    5333.24 &   57\\
Si I  &    5337.99 &   57\\
Si I  &    5338.21 &   27 \\
Fe I  &    5339.93 &   72\\
Fe I  &    5341.02 &   82\\
Mn I  &    5341.06 &   33\\
Cr I  &    5345    &   67\\
Cr I  &    5348    &   52\\
\hline                                                 
 Fe     &      5350-5500     &     $100\%$ used \\
\hline

Fe I   &   5353.37  &  33   \\
 Fe II &    5362.87 &   43\\
 Fe I  &    5364.87 &   62 \\
 Fe I  &    5365.40 &   42 \\
 Fe I  &    5367.47 &   68\\
 Fe I  &    5369.96 &   70\\
 Fe I  &    5371.44 &   14\\
 Fe I  &    5371.49 &  111 \\
 Fe I  &    5371.60 &   22\\
 Fe I  &    5383.37 &   77\\
 Fe I  &    5389.48 &   29 \\
 Fe I  &    5391.46 &   31\\
 Fe I  &    5393.17 &   65\\
Fe I   &   5397.13  & 105 \\
Fe I   &   5397.20  &  52 \\
Fe I   &   5400.50  &  49 \\
Fe I   &   5404.12  &  79 \\
Fe I   &   5404.15  &  75 \\
Si I   &   5404.66  &  44 \\
                Fe I   &   5405.77  & 107 \\
Cr I   &   5409.78  &  77 \\
Si I   &   5409.85  &  28 \\
                Fe I   &   5410.91  &  64  \\
Si I   &   5413.10  &  22 \\
 Fe I   &   5415.20  &  75 \\
 Fe I   &   5424.07  &  77 \\ 
 Fe I   &   5429.70  & 107 \\
 Fe I   &   5429.84  &  28 \\
 Fe I   &   5434.52  &  99 \\
 Fe I   &   5445.04  &  56 \\
 Fe I   &   5446.87  &  55 \\
 Fe I   &   5446.92  & 105 \\
 Fe I   &   5455.45  &  71  \\
 Fe I   &   5455.61  & 101  \\
 Fe I   &   5462.96  &  40 \\
 Fe I   &   5463.28  &  60 \\
 Fe I   &   5466.40  &  30 \\
 Fe I   &   5473.90  &  39  \\
 Fe I   &   5476.29  &  19 \\
 Fe I   &   5476.56  &  48 \\
Ni I   &   5476.90  &  86  \\
                Fe I   &   5487.74  &  43\\
                Fe I   &   5487.77  &  33\\
Si I   &   5488.98  &  45\\
Si I   &   5493.88  &  29\\
 Fe I   &   5497.52  &  79\\                   

 \hline                                       
 Fe      &     6590-6700     &       $100\%$ used \\   
 \hline
 Fe I   &   6592.91  &  72  \\
                 Fe I   &   6593.87  &  52\\
 Ni I   &   6643.63  &  44\\
                 Fe I   &   6663.44  &  44\\
                 Fe I   &   6677.98  &  82 \\
                                                                           
\hline                                                                           
 Fe      &     7440-7520     &       $37.5\%$ used \\
 \hline
 
Fe I   &   7445.75  &  76  \\ 
Cr I   &   7462.38  &  39\\
Fe II  &   7462.41  &  22\\
Fe I   &   7495.07  &  88\\
Fe I   &   7511.02  &  98\\
\hline                                                                                                                                        
 Fe      &     7775-7835     &       $66.7\%$ used \\
 \hline
Fe I   &   7780.56  &  74   \\
Fe I   &   7832.20  &  82 \\

\hline                                                                                                                                            
 Fe      &     8350-8490     &       $57.1\%$ used  \\
\hline
 Fe I   &   8365.63  &  29  \\
 Fe I   &   8387.77 &  133 \\
 Fe I   &   8468.41 &  101\\
 
\hline                                                       
Ca       &    3875-4000      &      $64.0\%$ used \\
\hline
Fe I    &  3920.26  &  94     \\
Fe I   &   3922.91 &   97    \\
Ca II  &   3933.66 &  73975                \\
Ca II  &   3968.47 &                       \\
\hline
Ca       &    4200-4250      &      $40.0\%$ used \\
\hline
Sr II   &  4215.52  & 503     \\
Fe I   &   4216.18  &  72  \\
Fe I   &   4217.55  &  56  \\
Fe I   &   4219.36  &  68  \\
Fe I   &   4222.21  &  69  \\
Fe I   &   4224.17  &  61  \\
Ca I   &   4226.73  & 370                   \\
Fe I   &   4227.43  &  75  \\
Fe II  &   4233.17  &  79  \\
Fe I   &   4233.60  &  76  \\ 
\hline                                                                                       
Ca     &      5575-5600       &     $92.0\%$ used   \\  
\hline
Fe I    &  5576.09  &  55     \\
Ca I   &   5581.96  &  29                  \\
Ca I   &   5588.75  &  80                  \\
Ca I   &   5590.11  &  29                  \\
Ca I   &   5594.46  &  70                  \\
Fe I   &   5598.29  &  31   \\
Ca I   &   5598.48  &  62                   \\
                                                                                             \hline
Ca     &      6100-6175        &    $82.7\%$ used \\
\hline
Ca I    &  6102.72  &  66                     \\
Ca I   &   6122.22 &   89                  \\
Si I   &   6155.13 &   38   \\
Ca I   &   6161.30 &   20                  \\
Ca I   &   6162.17 &  106                  \\
Ca I   &   6163.75 &   20                   \\
Ca I   &   6166.44 &   32                   \\
Ca I   &   6169.04 &   48                   \\
Ca I   &   6169.56 &   67              \\
\hline                                                                                               
Ca      &     6430-6500         &   $77.1\%$ used \\
\hline
Fe I   &   6430.84  &  78      \\
Ca I  &    6439.07 &  102                  \\
Ca I  &    6449.81 &   51                  \\
Ca I  &    6462.57 &   97                  \\
Ca I  &    8471.66 &   48                  \\
\hline                                                                                               
Ca      &     8450-8700        &    $84.0\%$ used \\
\hline
Ca II  &   8498.02 &  3100                     \\
Fe I   &   8514.07 &   93   \\
Ca II  &   8542.09 &  13266                \\
Ca II  &   8662.14 &  10091                \\
Fe I   &   8688.62 &  151  \\ 
\hline                                                                                             
Mg      &     5100-5250      &     $ 28.7\%$ used \\
\hline
Fe I   &   5139.25  &  75      \\
Fe I   &   5139.46  &  82  \\
MgH    &   5140     &                      \\
Fe I   &   5142.93  &  68  \\
Mg I   &   5167.32  & 119                  \\
Fe I   &   5167.49  & 103   \\
Fe I   &   5168.90  &  68   \\
Fe II  &   5169.03  & 223  \\
Fe I   &   5171.60  &  90  \\
Mg I   &   5172.68  & 148                 \\
Mg I   &   5183.60  & 190             \\
\hline\hline
\end{longtable}

}

\onllongtabL{2}{
\begin{landscape}
\begin{longtable}{lcccccccccccccccc}
\caption{Stellar parameters, chemical abundances, and distance estimates for our sample of SDSS/SEGUE stellar spectra. The identifier (ID) for each object is written as is common in SDSS, with the number of plate, mjd (modified julian date) and fiber associated to its observation. Full table available in electronic form\label{tbl-2}}\\
\hline
\hline
ID & b (rad) & l (rad)& $T_{\rm eff}$ (K) & $e_{T_{\rm eff}} (K)$ & $\log g$ [cgs] & $e_{\log g}$ [cgs] & [Fe/H] & $e_{[Fe/H]}$ & [Ca/H] & $e_{[Ca/H]}$ & [Mg/H] & $e_{[Mg/H]}$ & $d_{Sun}$ (kpc)& $e_{d_{Sun}}$/$d_{Sun}$ & r (kpc) & Z (kpc) \\
\hline
\endfirsthead
\caption{Continued.} \\
\hline
ID & b (rad) & l (rad)& $T_{\rm eff}$ (K) & $e_{T_{\rm eff}} (K)$ & $\log g$ [cgs] & $e_{\log g}$ [cgs] & [Fe/H] & $e_{[Fe/H]}$ & [Ca/H] & $e_{[Ca/H]}$ & [Mg/H] & $e_{[Mg/H]}$ & $d_{Sun}$ (kpc)& $e_{d_{Sun}}$/$d_{Sun}$ & r (kpc) & Z (kpc) \\
\hline
\endhead
\hline
\endfoot
\hline
\endlastfoot
0270-51909-0595         &      0.76         &      4.22         &      6252         &        32         &      3.16         &      0.24         &     -1.49         &      0.12         &     -1.17         &      0.04         &     -0.81         &      0.17         &     18.36         &      0.97         &     22.37         &     12.69        \\
0272-51941-0086         &      0.78         &      4.29         &      6083         &        48         &      3.49         &      0.28         &     -1.16         &      0.13         &     -0.70         &      0.06         &     -1.45         &      0.34         &     11.84         &      0.63         &     16.11         &      8.34        \\
0273-51957-0424         &      0.82         &      4.29         &      6273         &        49         &      2.97         &      0.36         &     -1.41         &      0.15         &     -1.21         &      0.06         &     -0.90         &      0.24         &     26.77         &      0.58         &     30.03         &     19.58        \\
0282-51630-0530         &      1.02         &      4.62         &      6086         &        28         &      3.34         &      0.23         &     -2.91         &      0.19         &     -2.51         &      0.04         &     -2.67         &      0.27         &     11.43         &      0.82         &     14.27         &      9.72        \\
0284-51662-0102         &      1.03         &      4.80         &      6034         &        43         &      2.79         &      0.32         &     -1.16         &      0.12         &     -1.29         &      0.06         &     -1.00         &      0.23         &     33.10         &      0.28         &     33.72         &     28.43        \\
0284-51662-0535         &      1.05         &      4.75         &      6138         &        48         &      3.83         &      0.30         &     -1.68         &      0.18         &     -1.17         &      0.07         &     -1.35         &      0.31         &      6.79         &      0.69         &     10.39         &      5.90        \\
0296-51984-0502         &      1.09         &      5.58         &      6257         &        29         &      2.54         &      0.21         &     -1.58         &      0.12         &     -1.22         &      0.04         &     -0.64         &      0.16         &     43.88         &      0.52         &     41.71         &     38.86        \\
0297-51663-0194         &      1.07         &      5.61         &      6249         &        26         &      2.80         &      0.20         &     -1.46         &      0.10         &     -1.05         &      0.03         &     -0.95         &      0.17         &     30.70         &      0.55         &     28.67         &     26.90        \\
0298-51955-0569         &      1.08         &      5.71         &      6081         &        44         &      3.69         &      0.25         &     -1.72         &      0.18         &     -0.91         &      0.06         &     -1.07         &      0.21         &     10.10         &      0.42         &     10.09         &      8.89        \\
0299-51671-0592         &      1.04         &      5.77         &      6192         &        55         &      3.20         &      0.46         &     -1.88         &      0.15         &     -4.24         &      0.35         &     -2.02         &      0.37         &     24.96         &      0.78         &     22.62         &     21.58        \\
0299-51671-0624         &      1.05         &      5.78         &      6056         &        25         &      2.96         &      0.18         &     -1.43         &      0.09         &     -1.17         &      0.03         &     -0.79         &      0.13         &     23.43         &      0.83         &     21.18         &     20.29        \\
0300-51666-0244         &      1.02         &      5.77         &      6101         &        46         &      2.68         &      0.41         &     -1.99         &      0.17         &     -1.83         &      0.05         &     -1.19         &      0.31         &     36.23         &      0.57         &     33.38         &     30.95        \\
0302-51688-0338         &      1.03         &      5.93         &      6288         &        58         &      3.05         &      0.45         &     -2.23         &      0.23         &     -2.07         &      0.08         &     -2.12         &      0.49         &     35.28         &      0.34         &     32.16         &     30.17        \\
0302-51688-0411         &      1.01         &      5.94         &      6123         &        40         &      3.83         &      0.27         &     -2.32         &      0.18         &     -1.88         &      0.05         &     -1.84         &      0.28         &      6.89         &      0.73         &      7.53         &      5.85        \\
0323-51615-0156         &      1.07         &      4.93         &      6280         &        32         &      3.82         &      0.21         &     -1.68         &      0.14         &     -1.02         &      0.03         &     -1.08         &      0.16         &      5.93         &      0.66         &      9.44         &      5.20        \\
0323-51615-0164         &      1.07         &      4.92         &      6221         &        32         &      3.62         &      0.21         &     -1.36         &      0.12         &     -0.91         &      0.04         &     -1.68         &      0.32         &      9.32         &      0.49         &     11.65         &      8.16        \\
0323-51615-0206         &      1.05         &      4.94         &      5973         &        32         &      3.77         &      0.21         &     -2.09         &      0.14         &     -1.49         &      0.04         &     -1.47         &      0.18         &      8.01         &      0.65         &     10.67         &      6.95        \\
0323-51615-0243         &      1.05         &      4.91         &      6124         &        37         &      2.61         &      0.31         &     -1.60         &      0.13         &     -1.51         &      0.04         &     -0.68         &      0.18         &     43.96         &      0.49         &     43.91         &     38.17        \\
0323-51615-0289         &      1.05         &      4.90         &      6159         &        28         &      2.51         &      0.23         &     -1.69         &      0.11         &     -1.58         &      0.04         &     -0.93         &      0.17         &     53.91         &      0.67         &     53.75         &     46.79        \\
0323-51615-0517         &      1.08         &      4.94         &      6128         &        20         &      2.71         &      0.15         &     -1.25         &      0.08         &     -1.11         &      0.02         &     -0.54         &      0.11         &     42.75         &      0.32         &     42.66         &     37.67        \\
0323-51615-0564         &      1.09         &      4.93         &      5897         &        19         &      2.81         &      0.13         &     -1.27         &      0.07         &     -1.07         &      0.01         &     -0.69         &      0.10         &     33.81         &      0.28         &     33.96         &     29.93        \\
0323-51615-0588         &      1.08         &      4.96         &      6223         &        25         &      2.75         &      0.20         &     -1.74         &      0.12         &     -1.03         &      0.02         &     -1.25         &      0.25         &     42.12         &      0.48         &     41.97         &     37.15        \\
0323-51615-0638         &      1.08         &      4.96         &      6115         &        27         &      3.79         &      0.16         &     -1.58         &      0.10         &     -1.06         &      0.03         &     -1.15         &      0.17         &      7.31         &      0.67         &     10.19         &      6.45        \\
0340-51691-0124         &      1.03         &      5.47         &      6164         &        32         &      2.73         &      0.26         &     -1.29         &      0.11         &     -1.29         &      0.04         &     -0.90         &      0.20         &     38.89         &      0.59         &     36.82         &     33.30        \\
0340-51691-0293         &      1.04         &      5.45         &      6215         &        42         &      2.76         &      0.35         &     -1.86         &      0.16         &     -1.71         &      0.05         &     -1.10         &      0.32         &     34.51         &      0.48         &     32.67         &     29.73        \\
0340-51691-0368         &      1.07         &      5.46         &      6163         &        49         &      3.09         &      0.38         &     -1.47         &      0.16         &     -1.30         &      0.06         &     -0.46         &      0.19         &     25.87         &      0.93         &     24.43         &     22.64        \\
0340-51691-0594         &      1.05         &      5.51         &      5952         &        35         &      3.77         &      0.22         &     -1.89         &      0.14         &     -1.36         &      0.05         &     -1.30         &      0.18         &      8.65         &      0.63         &      9.47         &      7.51        \\
0340-51990-0297         &      1.04         &      5.45         &      6209         &        36         &      2.83         &      0.29         &     -2.03         &      0.15         &     -1.81         &      0.04         &     -1.31         &      0.28         &     32.20         &      0.59         &     30.43         &     27.74        \\
0342-51691-0018         &      0.67         &      0.13         &      5941         &        37         &      2.64         &      0.28         &     -1.21         &      0.07         &     -1.29         &      0.04         &     -0.87         &      0.13         &     29.17         &      0.59         &     23.51         &     18.14        \\
0342-51691-0489         &      0.70         &      0.14         &      6053         &        25         &      2.56         &      0.19         &     -1.42         &      0.08         &     -1.17         &      0.03         &     -1.33         &      0.20         &     36.76         &      0.60         &     31.14         &     23.63        \\
0345-51690-0463         &      0.62         &      0.21         &      6088         &        44         &      3.11         &      0.36         &     -1.17         &      0.08         &     -2.93         &      0.06         &     -0.63         &      0.11         &     18.76         &      0.86         &     13.31         &     10.90        \\
0346-51693-0268         &      0.60         &      0.21         &      6124         &        16         &      2.71         &      0.12         &     -1.62         &      0.05         &     -1.39         &      0.01         &     -0.85         &      0.07         &     14.99         &      0.54         &      9.75         &      8.48        \\
0348-51696-0305         &      0.54         &      0.25         &      5890         &        51         &      2.59         &      0.40         &     -1.63         &      0.14         &     -1.43         &      0.06         &     -0.43         &      0.16         &     45.68         &      0.50         &     39.30         &     23.56        \\
0350-51691-0537         &      0.59         &      1.67         &      6181         &        29         &      3.19         &      0.22         &     -1.48         &      0.10         &     -1.18         &      0.03         &     -0.83         &      0.14         &     13.02         &      0.81         &     15.85         &      7.30        \\
0351-51695-0201         &      0.64         &      1.57         &      5815         &        54         &      3.64         &      0.35         &     -2.23         &      0.24         &     -1.61         &      0.07         &     -1.74         &      0.34         &     13.80         &      0.53         &     15.93         &      8.25        \\
0354-51792-0201         &      0.60         &      1.54         &      6298         &        22         &      3.04         &      0.15         &     -1.19         &      0.08         &     -1.03         &      0.02         &     -0.75         &      0.13         &     15.66         &      0.79         &     17.42         &      8.80        \\

...
\\                    
\end{longtable}
\end{landscape}
}

\onllongtabL{3}{
\begin{landscape}
\begin{longtable}{lcccccccccccccccc}
\caption{Similar data as in Table\ref{tbl-2} corresponding to our analyzed BOSS sample. Full table available in electronic form \label{tbl-3}}\\
\hline
\hline
ID & b (rad) & l (rad)& $T_{\rm eff}$ (K) & $e_{T_{\rm eff}} (K)$ & $\log g$ [cgs] & $e_{\log g}$ [cgs] & [Fe/H] & $e_{[Fe/H]}$ & [Ca/H] & $e_{[Ca/H]}$ & [Mg/H] & $e_{[Mg/H]}$ & $d_{Sun}$ (kpc)& $e_{d_{Sun}}$/$d_{Sun}$ & r (kpc) & Z (kpc) \\
\hline
\endfirsthead
\caption{Continued.} \\
\hline
ID & b (rad) & l (rad)& $T_{\rm eff}$ (K) & $e_{T_{\rm eff}}$ (K) & $\log g$ [cgs] & $e_{\log g}$ [cgs] & [Fe/H] & $e_{[Fe/H]}$ & [Ca/H] & $e_{[Ca/H]}$ & [Mg/H] & $e_{[Mg/H]}$ & $d_{Sun}$ (kpc)& $e_{d_{Sun}}$/$d_{Sun}$ & r (kpc) & Z (kpc) \\
\hline
\endhead
\hline
\endfoot
\hline
\endlastfoot
3587-55182-0432         &     -1.10         &      1.97         &      6079         &        24         &      3.79         &      0.15         &     -1.64         &      0.10         &     -1.26         &      0.03         &     -2.02         &      0.24         &      9.63         &      0.27         &     13.56         &     -8.56        \\
3588-55184-0102         &     -1.10         &      2.06         &      5992         &        14         &      3.76         &      0.09         &     -2.10         &      0.07         &     -1.60         &      0.01         &     -1.85         &      0.11         &      6.45         &      0.28         &     11.31         &     -5.75        \\
3606-55182-0147         &     -1.03         &      2.71         &      6043         &        15         &      3.74         &      0.10         &     -2.29         &      0.08         &     -1.95         &      0.02         &     -2.62         &      0.17         &      6.01         &      0.23         &     12.03         &     -5.16        \\
3615-55179-0098         &     -0.95         &      2.96         &      5908         &        17         &      3.63         &      0.09         &     -1.63         &      0.07         &     -1.06         &      0.01         &     -1.15         &      0.10         &     10.12         &      0.42         &     16.11         &     -8.21        \\
3615-55208-0100         &     -0.95         &      2.96         &      5940         &        15         &      3.82         &      0.08         &     -1.39         &      0.06         &     -1.06         &      0.01         &     -1.27         &      0.09         &      7.37         &      1.01         &     13.64         &     -5.98        \\
3615-55445-0096         &     -0.95         &      2.96         &      5952         &        20         &      3.73         &      0.11         &     -1.52         &      0.08         &     -1.06         &      0.02         &     -1.23         &      0.13         &      9.07         &      0.34         &     15.16         &     -7.36        \\
3647-55476-0178         &     -0.96         &      2.96         &      5894         &        24         &      3.87         &      0.14         &     -1.77         &      0.09         &     -1.44         &      0.03         &     -1.53         &      0.13         &      7.42         &      1.09         &     13.65         &     -6.06        \\
3647-55827-0166         &     -0.96         &      2.96         &      6044         &        23         &      3.85         &      0.13         &     -1.68         &      0.09         &     -1.31         &      0.03         &     -1.57         &      0.16         &      7.18         &      0.96         &     13.44         &     -5.87        \\
3650-55244-0114         &     -0.92         &      3.02         &      6028         &        22         &      3.76         &      0.13         &     -1.77         &      0.09         &     -1.14         &      0.04         &     -1.57         &      0.16         &      9.29         &      0.34         &     15.48         &     -7.39        \\
3650-55244-0993         &     -0.90         &      3.01         &      6021         &        22         &      3.76         &      0.14         &     -1.65         &      0.10         &     -1.23         &      0.03         &     -1.45         &      0.17         &      7.04         &      0.32         &     13.54         &     -5.50        \\
3658-55205-0478         &      0.42         &      3.10         &      6256         &        56         &      3.69         &      0.38         &     -2.65         &      0.34         &     -1.97         &      0.08         &     -1.85         &      0.43         &     16.05         &      0.61         &     23.58         &      6.51        \\
3659-55181-0934         &      0.46         &      3.06         &      6129         &        32         &      3.60         &      0.20         &     -1.42         &      0.12         &     -1.03         &      0.03         &     -1.75         &      0.32         &     14.19         &      0.45         &     21.64         &      6.30        \\
3659-55181-0972         &      0.46         &      3.07         &      6142         &        37         &      3.77         &      0.22         &     -1.07         &      0.14         &     -0.90         &      0.04         &     -0.84         &      0.20         &     11.62         &      0.73         &     19.11         &      5.17        \\
3660-55209-0192         &      0.46         &      3.04         &      6082         &        40         &      3.61         &      0.25         &     -1.86         &      0.17         &     -1.30         &      0.05         &     -1.78         &      0.36         &     17.21         &      0.33         &     24.61         &      7.64        \\
3660-55209-0792         &      0.46         &      3.02         &      6070         &        16         &      2.77         &      0.12         &     -2.37         &      0.10         &     -1.84         &      0.02         &     -2.08         &      0.19         &     17.81         &      0.29         &     25.18         &      7.97        \\
3663-55176-0600         &      0.46         &      3.00         &      5852         &        47         &      3.75         &      0.27         &     -1.97         &      0.18         &     -1.31         &      0.06         &     -1.68         &      0.27         &     16.64         &      0.36         &     24.03         &      7.31        \\
3664-55245-0282         &      0.44         &      3.15         &      5939         &        37         &      3.76         &      0.21         &     -1.93         &      0.17         &     -1.26         &      0.05         &     -1.63         &      0.26         &     13.85         &      0.33         &     21.36         &      5.91        \\
3664-55245-0358         &      0.44         &      3.13         &      5890         &        31         &      3.67         &      0.17         &     -1.58         &      0.12         &     -1.21         &      0.04         &     -1.39         &      0.19         &     12.26         &      0.34         &     19.79         &      5.24        \\
3664-55245-0662         &      0.45         &      3.10         &      6142         &        45         &      2.96         &      0.34         &     -2.67         &      0.30         &     -1.61         &      0.05         &     -2.19         &      0.50         &     48.59         &      0.31         &     55.90         &     21.14        \\
3665-55247-0562         &      0.47         &      2.98         &      6043         &        39         &      3.55         &      0.23         &     -1.53         &      0.14         &     -1.11         &      0.05         &     -1.04         &      0.17         &     16.61         &      0.56         &     23.95         &      7.52        \\
3665-55247-0622         &      0.48         &      2.96         &      5988         &        35         &      3.76         &      0.20         &     -1.92         &      0.15         &     -1.30         &      0.04         &     -1.83         &      0.27         &     14.68         &      0.30         &     22.00         &      6.79        \\
3665-55247-0930         &      0.49         &      2.99         &      6003         &        24         &      3.44         &      0.15         &     -1.37         &      0.09         &     -1.25         &      0.03         &     -1.25         &      0.18         &     12.92         &      0.36         &     20.26         &      6.14        \\
3666-55185-0548         &      0.45         &      3.08         &      5946         &        36         &      3.49         &      0.21         &     -1.53         &      0.12         &     -1.11         &      0.05         &     -1.42         &      0.23         &     15.95         &      0.37         &     23.40         &      6.94        \\
3668-55478-0505         &      0.48         &      2.95         &      6241         &        45         &      3.71         &      0.31         &     -2.02         &      0.21         &     -1.71         &      0.06         &     -2.00         &      0.44         &     14.00         &      0.48         &     21.33         &      6.49        \\
3668-55478-0590         &      0.50         &      2.93         &      6116         &        42         &      3.84         &      0.28         &     -2.84         &      0.28         &     -1.95         &      0.06         &     -2.27         &      0.35         &     11.21         &      0.98         &     18.54         &      5.35        \\
3668-55478-0716         &      0.50         &      2.94         &      5982         &        30         &      3.49         &      0.20         &     -1.91         &      0.12         &     -1.36         &      0.04         &     -1.53         &      0.20         &     20.23         &      0.67         &     27.41         &      9.70        \\
3668-55478-0790         &      0.51         &      2.93         &      5920         &        37         &      3.46         &      0.25         &     -2.07         &      0.15         &     -1.53         &      0.04         &     -1.60         &      0.23         &     19.92         &      0.64         &     27.08         &      9.64        \\
3670-55480-0498         &      0.44         &      3.15         &      5990         &        41         &      3.75         &      0.25         &     -1.84         &      0.17         &     -1.21         &      0.06         &     -1.53         &      0.28         &     13.38         &      0.31         &     20.90         &      5.71        \\
3670-55480-0704         &      0.47         &      3.15         &      5827         &        52         &      3.68         &      0.38         &     -3.19         &      0.41         &     -2.44         &      0.07         &     -2.60         &      0.43         &     19.98         &      0.44         &     27.37         &      8.96        \\
3671-55483-0258         &      0.48         &      3.06         &      6061         &        39         &      3.71         &      0.26         &     -2.33         &      0.20         &     -2.20         &      0.05         &     -2.89         &      0.47         &     13.26         &      0.26         &     20.66         &      6.17        \\
3673-55178-0520         &      0.48         &      3.01         &      6087         &        43         &      3.81         &      0.24         &     -1.77         &      0.17         &     -1.01         &      0.05         &     -1.91         &      0.37         &     12.76         &      0.38         &     20.15         &      5.91        \\
3673-55178-0742         &      0.49         &      2.99         &      6021         &        29         &      3.65         &      0.17         &     -1.77         &      0.12         &     -1.21         &      0.04         &     -1.19         &      0.18         &     11.09         &      0.39         &     18.48         &      5.27        \\
3676-55186-0272         &      0.50         &      3.07         &      6159         &        38         &      3.81         &      0.23         &     -1.32         &      0.14         &     -1.17         &      0.05         &     -1.35         &      0.27         &     10.71         &      0.78         &     18.13         &      5.15        \\
3683-55178-0688         &      0.53         &      3.08         &      5972         &        42         &      3.74         &      0.24         &     -1.79         &      0.16         &     -1.18         &      0.06         &     -1.37         &      0.23         &     15.17         &      0.33         &     22.43         &      7.66        \\
3683-55245-0682         &      0.53         &      3.08         &      5988         &        36         &      3.78         &      0.20         &     -2.12         &      0.18         &     -1.25         &      0.05         &     -1.62         &      0.25         &     14.41         &      0.29         &     21.69         &      7.27        \\
3691-55274-0886         &      0.59         &      3.04         &      6146         &        46         &      3.74         &      0.30         &     -2.35         &      0.24         &     -1.76         &      0.05         &     -1.46         &      0.29         &     13.84         &      0.34         &     20.95         &      7.67        \\
...
\\                    
\end{longtable}
\end{landscape}
}

\end{document}